\documentclass[aps,pra,singlecolumn,superscriptaddress,floatfix,
nofootinbib,showpacs,longbibliography,notitlepage]{revtex4-1}

\usepackage[utf8]{inputenc}  
\usepackage[T1]{fontenc}     
\usepackage[british]{babel}  
\usepackage[sc,osf]{mathpazo}
\usepackage[scaled=0.86]{berasans}  
\usepackage[colorlinks=true, citecolor=blue, urlcolor=blue]{hyperref}  
\usepackage{graphicx} 
\usepackage[babel]{microtype}  
\usepackage{amsmath,amssymb,amsthm,bm,amsfonts,mathrsfs,bbm} 

\usepackage{xspace}  
\usepackage{pgf,tikz}
\usepackage{xcolor}
\usepackage{multirow}
\usepackage{array}
\usepackage{bigstrut}
\usepackage{braket}
\usepackage{color}
\usepackage{natbib}
\usepackage{multirow}
\usepackage{mathtools}
\usepackage{float}
\usepackage[caption = false]{subfig}
\usepackage{xcolor,colortbl}
\usepackage{color}
\newcommand{\Tr}{\operatorname{Tr}}

\newcommand{\be}{\begin{equation}}
\newcommand{\ee}{\end{equation}}
\newcommand{\ba}{\begin{eqnarray}}
\newcommand{\ea}{\end{eqnarray}}

\newtheorem{lemma}{Lemma}

\newcommand{\N}{\mathbb{N}}

\newcommand{\R}{\mathbb{R}}
\newcommand{\C}{\mathbb{C}}

\newcommand{\spc}[1]{\mathcal{#1}}

\def\>{\rangle}
\def\<{\langle}
\def\kk{\>\!\>}
\def\bb{\<\!\<}
\newcommand{\st}[1]{\mathbf{#1}}
\newcommand{\bs}[1]{\boldsymbol{#1}}     

\newcommand{\map}[1]{\mathcal{#1}}

\newtheorem{theo}{Theorem}

\newtheorem{cor}{Corollary}

\newcommand{\Proof}{{\bf Proof. \,}}

\begin{document}

\title{Indefinite causal order enables perfect quantum communication with  zero capacity channels}

\author{Giulio Chiribella}
\affiliation{QICI Quantum Information and Computation Initiative, Department of Computer Science, The University of Hong Kong, Pokfulam Road, Hong Kong.}
\affiliation{Department of Computer Science, University of Oxford, Wolfson Building, Parks Road, United Kingdom.}
\affiliation{Perimeter Institute for Theoretical Physics, Waterloo N2L 2Y5, Canada}

\author{Manik Banik}
\affiliation{School of Physics, IISER Thiruvanathapuram, Vithura, Kerala 695551, India.}

\author{Some Sankar Bhattacharya}
\affiliation{QICI Quantum Information and Computation Initiative, Department of Computer Science, The University of Hong Kong, Pokfulam Road, Hong Kong.} 

\author{Tamal Guha}
\affiliation{Physics and Applied Mathematics Unit, Indian Statistical Institute, 203 B.T. Road, Kolkata-700108, India.}

\author{Mir Alimuddin}
\affiliation{Physics and Applied Mathematics Unit, Indian Statistical Institute, 203 B.T. Road, Kolkata-700108, India.}

\author{Arup Roy}
\affiliation{Department of Physics, A B N Seal College Cooch Behar, West Bengal 736101, India.}

\author{Sutapa Saha}
\affiliation{Physics and Applied Mathematics Unit, Indian Statistical Institute, 203 B.T. Road, Kolkata-700108, India.}

\author{Sristy Agrawal}
\affiliation{Department of Physics and Center for Theory of Quantum Matter, University of Colorado, Boulder, Colorado 80309, USA.}
\affiliation{National Institute of Standards and Technology, Boulder, Colorado 80305, USA.}

\author{Guruprasad Kar}
\affiliation{Physics and Applied Mathematics Unit, Indian Statistical Institute, 203 B.T. Road, Kolkata-700108, India.}

\begin{abstract}
Quantum mechanics is compatible with scenarios where the relative order between two events can be indefinite. Here we show that two independent instances  of  a noisy process can behave as a perfect quantum communication channel when used  in a coherent superposition of two alternative orders. 
 This phenomenon occurs even if the original process  has zero capacity to transmit quantum information. 
 In contrast, perfect quantum communication does not occur when the message is sent directly from the sender to the receiver through a superposition of alternative paths,  with an independent noise process acting on each path. 
The possibility of perfect quantum communication through independent noisy channels highlights a fundamental difference between the superposition of orders in time and the superposition of paths in space. 
 \end{abstract}



\maketitle

\section{Introduction}

The  framework of information theory  was established in the  seminal work of Claude Shannon \cite{Shannon48}, who laid the foundations of our current communication  technology.   In his work, Shannon modelled the devices used to store and transfer information as classical systems, whose  internal state can in principle be determined without errors, and whose arrangement in space and time is always well-defined.  At the fundamental level, however, physical systems obey  the laws of quantum mechanics, which in principle can be exploited to achieve communication tasks that are impossible in classical physics \cite{Bennett85, Ekert91, Bennett92}.   
The ability of quantum channels to transmit information has been quantified by various types of quantum capacities, such as the classical capacity \cite{Holevo98,Schumacher97}, the quantum  capacity \cite{Lloyd97,Shor02,Devetak05}, and the entanglement-assisted capacity   \cite{Bennett96,Bennett99}. By now, the theory of communication with quantum systems has become   a  throughly  developed discipline, known as quantum Shannon theory \cite{Chuang00,Wilde13}.

The standard model of communication in quantum Shannon theory  generally assumes that the available communication channels are used in a definite  configuration. In principle, however, quantum theory is compatible with   scenarios where the configuration of the communication channels is in a quantum superposition.   For example, a photon could travel through a superposition of different paths between the sender and receiver  \cite{Aharonov90, Oi03, aaberg2004subspace}, and interference between the noise processes on different paths could offer  an  opportunity to filter out some of the noise affecting the transmission \cite{Gisin05}.   More recently, it has been observed that the superposition of channel configurations can also involve the order of the channels in time, in a scenario known as the  {\em quantum SWITCH} \cite{chiribella2009beyond,Chiribella13}. 
    In the quantum SWITCH, the relative order of  two channels is controlled by a qubit, and superpositions in the state of such qubit lead to indefinite causal order.  The indefiniteness of the order is sometimes called  causal non-separability \cite{Oreshkov12,Araujo15,Oreshkov16}.  
    
  In recent years, the applications of the quantum SWITCH and other causally non-separable processes have attracted increasing interest, leading to the discovery of quantum advantages in various tasks, such as testing  properties of quantum channels \cite{Chiribella12, Araujo14}, winning non-causal games \cite{Oreshkov12}, reducing quantum communication complexity \cite{Guerin16}, boosting the precision of quantum metrology \cite{zhao2020quantum}, and achieving thermodynamic advantages \cite{Guha20}. Experimental investigations of the quantum SWITCH   have been recently proposed in various photonic setups \cite{Procopio15, Rubino17, Goswami18(1),wei2019experimental,guo2020experimental,goswami2020experiments,rubino2020experimental}. The quantum SWICH also admits more exotic realizations, which could take place in new physical regimes involving quantum superpositions of spacetimes \cite{Zych17} or closed timelike curves \cite{chiribella2009beyond,Chiribella13}. 

 The extension of quantum Shannon theory to scenarios involving the superposition of causal orders  has been recently initiated by Ebler, Salek, and one of the authors \cite{Ebler18,Salek18}. In these works, the authors established a number of advantages  with respect to the standard communication model of quantum Shannon theory, where channels are arranged in a definite configuration, and no additional side channels are used \cite{Hler20}.     Recently, some of the Shannon theoretic advantages of the quantum SWITCH  have been demonstrated  experimentally  in photonic setups \cite{Goswami18,guo2020experimental,rubino2020experimental}.   A natural question is whether  these advantages are an exclusive feature of the superposition of orders, or whether instead they could be reproduced by the  superposition of paths originally considered in Refs. \cite{Aharonov90, Oi03, aaberg2004subspace}. Recently, Abbott {\em et al} \cite{Abbott2018} argued for the latter, showing  that some of the advantages of the quantum SWITCH can be reproduced  in a scenario where multiple independent channels are put in parallel between the sender and receiver,  and the message is sent through them along a superposition of paths.  But can all the advantages of the quantum SWITCH  be reproduced in this way?

        Here we answer the question in the negative, showing that the combination of two independent channels in an indefinite order leads to a phenomenon that cannot be achieved through the combination of independent channels on alternative paths between the sender and the receiver.    Specifically, we show that an entanglement-breaking channel, which in normal conditions cannot send any quantum information,  can become a perfect  quantum communication channel when two independent uses of it are combined by the quantum SWITCH.
       In contrast, we prove that perfect quantum communication cannot take place when the two independent uses of the channel are placed on  two alternative paths between the sender and the receiver, letting the message travel on a coherent superposition of these paths.  More generally, we show that no superposition of any  finite number of independent noisy channels can lead to a complete noise removal.  Our result proves that any communication model that reproduces all the advantages of the quantum SWITCH  through the superposition of paths between the sender and the receiver  must necessarily feature correlations between the processes occurring on the different paths  \cite{chiribella2019quantum,Hler20, kristjansson2020single}.

The quantum communication advantage established here is also interesting   as an extreme form of activation of the quantum capacity. In our example,  two  channels with zero quantum capacity are combined into a channel that has not only positive capacity, but also {\em maximal} capacity for the given input size.  We  characterize the set of channels that give rise to such extreme form of activation, showing that our example is unique up to changes of basis.

In Section \ref{switch} we briefly review the quantum SWITCH and its application to quantum Shannon theory.  In Section \ref{perfect} we present  our example of perfect activation of channels with zero quantum capacity, and we show that, for qubit, our example is essentially unique.  In Section \ref{path} we prove that  perfect activation  cannot be achieved through the superposition of independent noisy channels.     In Section \ref{implications} we discuss the implication of the theorem proved in Section \ref{path}.  The conclusions of the paper are given in Section \ref{con}.

\section{Quantum SWITCH}\label{switch}

A general quantum process, transforming an input system  $A$  into an output system $B$, is described by a quantum channel,  namely a linear, completely positive, trace-preserving map from ${\sf L}(\spc H_A)$ to ${\sf L}(\spc H_B)$, where $\spc H_A$ and $\spc H_B$ denote the Hilbert spaces of systems $A$ and $B$, respectively,   and ${\sf L}  (\spc H)$ denotes the space of linear operators on a generic Hilbert space $\spc H$.     The set of density operators on the Hilbert space $\spc H$ will be denoted as ${\sf D} (   \spc  H)$. The action of a generic quantum channel $\map E$ on an input state $\rho\in  {\sf D}  (\spc H_A)$ can be expressed in the Kraus representation  as $\map E (\rho)  =  \sum_i E_i \rho E_i^\dag$, where the Kraus operators $E_i  :  \spc H_A \to \spc H_B$ are linear operators satisfying $\sum_i E_i^\dag E_i  =  I_A$, $I_A$ being the identity operator on $\spc H_A$. 

Two communication channels $\map E$ and $\map F$  can be combined in different configurations, either by Nature itself or by a communication provider that sets up  the communication network between the sender and the receiver  \cite{Hler20}.   Classically, the channels $\map E$ and $\map F$ can be combined in a variety of well-defined configurations. For example, they can be combined in parallel, giving rise to the product channel $\map E \otimes \map F$, or  in a sequence (if  their inputs and outputs match), giving rise either to the channel $\map E \circ \map F$ or to the channel $\map F \circ  \map E$.  The parallel configuration corresponds to the scenario where the two channels are used by a sender to communicate directly to a receiver. The sequential configuration corresponds to  the scenario where the information  travels through two causally connected regions before reaching the receiver.     More generally, one could think of a sequential composition where a third process $\map R$ takes place in between  $\map E$ and $\map F$. Here  $\map R$ could be  an operation performed at an intermediate station  placed between the sender and the receiver.

  In principle, quantum theory allows for scenarios where two processes, $\map E$ and $\map F$,  are combined in a quantum superposition of two alternative orders, via a higher-order operation called the quantum SWITCH \cite{chiribella2009beyond,Chiribella13}. 
For simplicity, consider the case of two processes $\map E$ and $\map F$ transforming system $A$ to system $B$, with $\spc H_A  
   \simeq \spc H_B$.       The new quantum channel  $\map S (\map E, \map F)$ resulting from the combination of  $\map E$ and $\map F$ in an order controlled by a control qubit  $C$ is described by the   Kraus operators  
  \begin{equation}\label{eq1}
S_{ij}=  E_i  F_j \otimes \ket{0}\bra{0}_{C}+  F_j E_i \otimes \ket{1}\bra{1}_{ C},
\end{equation}  
where $\{E_i\}$ and $\{  F_j\}$ are  the Kraus operators of the channels $\map E$ and $\map F$, respectively, and $\{  |0\>_{ C},  |1\>_{C}\}$ are orthogonal states of system $\rm C$.  Note that the definition of the channel $\map S(\map E,\map F)$ is independent of the choice of Kraus representation used for $\map E$ and $\map F$ \cite{Chiribella13}.  Mathematically, the quantum SWITCH is a {\em quantum supermap}  \cite{chiribella2008transforming,chiribella2009theoretical,Chiribella13}, transforming a pair of quantum channels $\map E$ and $\map F$ into the new quantum channel $\map S (\map E,\map F)$.

      \begin{figure}
	\includegraphics[scale=0.85]{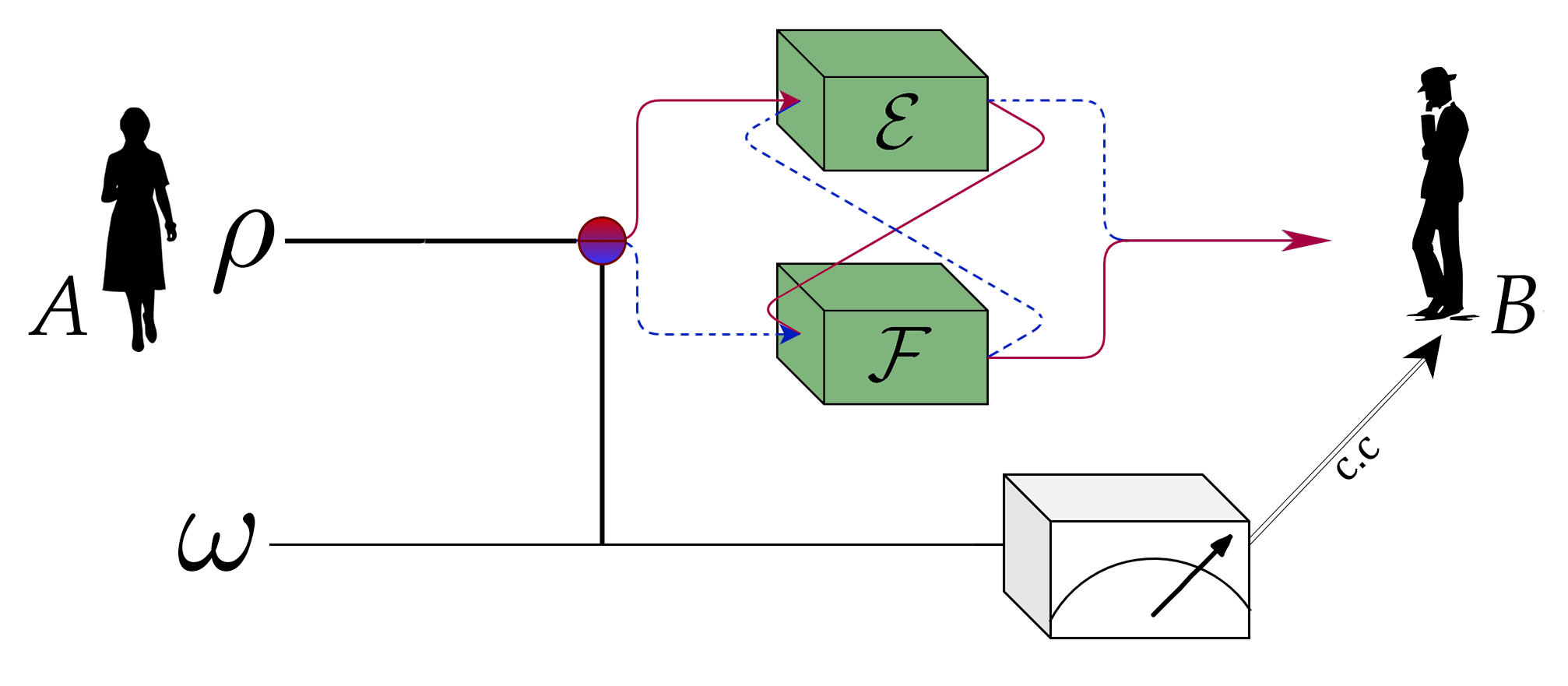}
	\caption{(Color on-line) {\bf Communication through the quantum SWITCH.} A communication provider places two communication channels $\mathcal{E}$ and $\mathcal{F}$ in a coherent superposition of two alternative causal orders, either with channel  $\mathcal{E}$ acting in the causal past of channel  $\mathcal{F}$ or the other way round. The order of the two channels is controlled by  a qubit  (bottom wire in the figure), initialized in the state $\omega$ by the communication provider.  A sender (A) sends  a quantum state 
	$\rho$ to a receiver (B) through the channels $\map E$ and $\map F$.  To assist the receiver,  the communication provider measures the  control qubit and communicates the  outcome via a classical transmission line (doubled line in the figure).   }
	\label{fig1}
\end{figure}

In a communication scenario, the quantum SWITCH supermap can be interpreted as an operation performed by a communication provider that sets up the communication network between the sender and the receiver \cite{Hler20}. In this model, the control system can be accessed by the communication provider, but remains inaccessible to the sender, who can only   encode information in the target system.    The same model is generally applied to the superposition of paths, where the path degree of freedom is not used to directly encode information, but only to assist the communication \cite{Gisin05,Abbott2018,chiribella2019quantum, Hler20}.     In both communication models, the  control (or path) system  is  set to a fixed state $\omega$,   viewed as a parameter of the communication network between the sender and the receiver.  In the case of the quantum SWITCH,  the effective channel between sender and receiver is the channel $\map S_\omega  (  \map E, \map F)$ defined by
\begin{equation}\label{eq2}
\mathcal{S}_{\omega}(\map E,\map F) (\rho):=\sum_{i,j} \, S_{ij}(\rho\otimes\omega)S_{ij}^{\dagger} \,.
\end{equation} 
For $\map E  =  \map  F$, the channel $\map S_\omega (\map E, \map E)$ has the simple form  
\begin{align}\label{switchsame}  
\mathcal{S}_{\omega}(\map E,\map E) (\rho)     =     \frac 1 4  \,   \left(    \sum_{i,j}   \{  E_i,  E_j\} \rho  \{  E_i,  E_j\}^\dag   \otimes \omega   +      [  E_i,  E_j] \rho  [ E_i,  E_j]^\dag   \otimes Z\omega Z \, \right) \, ,   
\end{align} 
where   $\{  E_i,E_j\} :  = E_iE_j +  E_jE_i$ and $[E_i,E_j] :  = E_iE_j-  E_jE_i$ are the anticommutator and the commutator, respectively.   Note that there is no entanglement between the target system and the control system at the output state of  channel $\map S_\omega  (\map E,\map E)$.

 In the following we  assume that   the  communication provider measures the control system and communicates the outcome to the receiver through a classical transmission line, as illustrated in Figure \ref{fig1}.     This  setting is similar to  the setting of  quantum communication with classical assistance from the environment \cite{Greg03, Smolin05}. An important difference, however, is that  we do not assume that the whole environment is accessible: the only part of the environment that needs to be accessible to the communication provider is the two-dimensional system  responsible for the order of the channels $\map E$ and $\map F$.

\section{Perfect activation of the quantum capacity}\label{perfect}
The relevant quantity of the transmission of quantum information  is  the coherent information \cite{Schumacher96}, defined as  \begin{align}
I_{\rm c}(\map E):=\max_{\rho \in   {\sf D} (   \spc  H_A  \otimes \spc H_A)}    I_{\rm c}( A  \>  B)_{(\map I_A \otimes  \map E)  (\rho )} \,,
\end{align} where $I_{\rm c}(A\>   B)_{\sigma}   :=   S(  \sigma_{B})  -  S(\sigma_{AB})$ is the coherent information of  a generic bipartite  state $\sigma   \equiv \sigma_{AB}   \in {\sf D} (\spc H_A \otimes \spc  H_B )$,  $\sigma_B$ is the marginal state $\sigma_B  := \Tr_A  [\sigma_{AB}]$, and $S(\tau)  :=  -  \Tr[\tau  \log \tau]$ is the von Neumann entropy of a generic quantum state $\tau$, with the logarithm taken in base 2.    When the state $\sigma$ is of the separable form $\sigma  = \sum_i  \,q_i\,  \sigma_{A, i}  \otimes \sigma_{B,i}$, one has   $I (A\>   B)_{\sigma}\le 0$, with the equality if and only if system $A$ is in a pure state.   This implies that entanglement-breaking channels, which transform every state into a separable state, have zero coherent information.        

When the channel $\map E$ is used in parallel for an asymptotically large number of times, its ability to transmit quantum information is measured by the  quantum capacity $Q   (\map E)$, which can be computed in terms of the coherent information as $Q(\map E)  =   \lim_{n\to \infty}  I_{\rm c}(\map E^{\otimes n})/n$ \cite{Lloyd97,Shor02,Devetak05}.   For an entanglement-breaking channel $\map E$,  the coherent information $ I_{\rm c}(\map E^{\otimes n})$ is zero for every $n$, and therefore the quantum capacity is zero \cite{Holevo99,Holevo08}.

We now show that the combination of two entanglement-breaking channels in the quantum SWITCH can generate a perfect quantum communication channel.  Our example involves two Pauli channels, that is, two qubit channels  $\map E_{\vec{p}}$  with  a Kraus decomposition of the form $\map E_{\vec{p}}(\rho)=  p_0  \, \rho  +   p_1\,  X\rho X+  p_2\, Y \rho Y  +  p_3\,  Z\rho Z$, where $\vec{p}\equiv(p_0,p_1,p_2,p_3)$ is a probability vector, and $(X, Y, Z)$ are the three Pauli matrices. 

Suppose that two uses of the same Pauli channel $\map E_{\vec{p}}$ are combined in the quantum SWITCH.  The action of the resulting channel can be obtained from Equation (\ref{switchsame}),  which yields 
\begin{widetext}
	\begin{align}
 \nonumber \map S_\omega (\map E_{\vec p},  \map E_{\vec p})  (\rho) =  q_+   \, \map C_+  (\rho) \otimes \omega_+  +   q_-  \,  \map C_-  (\rho) \otimes \omega_-  \, , \qquad      q_-  =  2(p_1p_2  +  p_2 p_3  +   p_3 p_1) \, , \qquad  q_+  = 1- q_-  &\\
   \nonumber \map C_+  (\rho)    =    \frac {\left(  p_0^2+p_1^2+p_2^2+p_3^2 \right)      \,  \rho    +  2p_0   \, \left(p_1 \,  X \rho  X  + p_2 \,  Y \rho Y + p_3  Z \rho Z \right)} {q_+}  &  \\
\label{pauliswitch}	\map C_-   (\rho)     =  \frac{  2 p_1 p_2   \,  Z  \rho Z  +2 p_2 p_3\,  X \rho X+  2 p_1 p_3\,  Y\rho Y }{q_-} \, ,   \qquad \omega_+  =  \omega \, , \qquad  \omega_-  = Z \omega  Z \, .  &
	\end{align}
\end{widetext}  
  
For $\omega  = |+\>\<+|$,   the final states of the control system are the orthogonal states $|+\>$ and $|-\>$, with $|\pm\>  =  (|0\>  \pm  |1\> )/\sqrt 2$.   In other words, the output of the channel  $\map S_{|+\>\<+|} (\map E_{\vec p},  \map E_{\vec p})$ exhibits perfect classical correlations between the evolution of the target system  and two orthogonal states of the control system.  

A measurement on the control system can then  separate the two  channels $\map C_+$ and $\map C_-$  in Equation   (\ref{pauliswitch}).        By measuring the environment and communicating the outcome to the receiver, a communication provider can improve the quality of the transmission, giving the receiver the opportunity to decode  the channels  $\map C_+$ and $\map C_-$ separately. 
 Now, the key point is that the channels $\map C_+$ and $\map C_-$  can be noiseless even if the original channel $\map E_{\vec p}$ was noisy: 
 \begin{itemize}
 \item  if $p_0$ is zero,  then channel $\map C_+$ is the identity, 
 \item  if  one of the three probabilities $p_1,p_2$, or $p_3$ is zero, then channel $\map C_-$ is unitary. 
 \end{itemize}    
   When both conditions are satisfied,  the quantum channel $\map S_{|+\>\<+|}  (\map E_{\vec p}, \map E_{\vec p})$ enables  a perfect, deterministic transmission of a qubit from the sender to the receiver. This is the case for the channel $\map E_{XY}  (\rho)   =    1/2 \,  ( X\rho X  +   Y  \rho Y) $. 
  
 The channel $\map E_{XY}  (\rho)$ exhibits   an extreme example of activation of the quantum capacity.    It is   entanglement-breaking, because its action can be equivalently expressed in the measure-and-reprepare form $\map E_{XY} (\rho)  =  |1\>\<1  |  \,  \<0| \rho  |0\>  +  |0\>\<0|  \,  \<1|\rho  |1\>$.    Hence, $\map E_{XY}$ has zero quantum capacity: in the standard communication model of quantum Shannon theory it cannot transmit any quantum information, even if  used infinitely many times in parallel or in sequence. 
In contrast,  the quantum channel $\map S_{|+\>\<+|} (\map E_{XY}, \map E_{XY})$  has unit capacity, which is the maximum capacity one could possibly obtain with a qubit input.   Recently, the extreme activation offered by the quantum SWITCH was experimentally observed in \cite{guo2020experimental}, up to a small error due  to the  unavoidable imperfections affecting any realistic setup.

Physically, one can ask which resources are responsible for the  activation of the quantum capacity shown in our example.   From the point of view of the communication provider, who sets up the communication network between sender and receiver, the resource is the ability to coherently control the order of two channels, and the ability to perform a measurement in  the basis $\{|+\>,  |-\>\}$, whose vectors are coherent superpositions of the vectors $\{|0\>, |1\>\}$ controlling the choice of orders.  If the the control qubit were prepared in an incoherent mixture of the states $\{|0\>, |1\>\}$, or if measurement were performed in the incoherent basis   $\{|0\>, |1\>\}$, then the evolution of the target system would be described by the entanglement-breaking channel $\map E_{XY}^2$, and no advantage could take place.  Hence, a key resource in the protocol is quantum coherence  \cite{streltsov2017colloquium} in the  qubit controlling the causal order.  
From the point of view of the sender and receiver, who don't have direct access to the control system,  the key resource is the correlation between the evolution of the target system, and the classical information received from the communication provider.  In this respect, our example can be seen as a special instance  of quantum communication with classical assistance from the environment \cite{Greg03, Smolin05}. 
 
At this point, it is natural to ask which quantum channels  exhibit the extreme activation phenomenon showed in our example.   Interestingly, we find out that our example is essentially unique. 
First of all,  we show  that every qubit channel $\map E$  satisfying the conditions $Q(\map E)  = 0$ and $Q (\map S_\omega  (\map E,\map E))  = 1$  must be unitarily equivalent to the channel $\map E_{XY}$. 
Since the quantum capacity  $Q (\map S_\omega  (\map E,\map E))  = 1$   quantifies the amount of information that can be decoded  with full access to the output of channel $\map S_\omega (\map E,\map E)$, our result covers in particular the case where the control is measured and the outcome is communicated to the receiver: if unit capacity is to be achieved at all,  then the channel $\map E$ must be unitarily equivalent to $\map E_{XY}$. 

To obtain the above result,   we first characterize the qubit channels that achieve unit capacity when inserted into the quantum SWITCH.  

\begin{theo}\label{theo:qubit}
 	For a qubit channel $\map E$, the unit-capacity condition $Q (\map S_{\omega}  (\map E, \map E))  =1$  is satisfied if and only if one of the following conditions is satisfied 
 	\begin{enumerate}
 		\item the channel $\map E$ is unitary 
 		\item  the control qubit is in a pure state $|\gamma\>$ with $|\<0|\gamma\>|  =|  \<1|\gamma\>|$   and the channel $\map E$ is of the random-unitary form $\map E (\rho)   =   q \,  (UXU^\dag)    \, \rho  \, (   UXU^\dag)  +  (1-q)   \,  (UYU^\dag) \,  \rho  \, (UYU^\dag)$, where $q   \in   [0,1]$ is a probability, $U$ is an arbitrary unitary gate, and $X$ and $Y$ are Pauli matrices.
 	\end{enumerate}
 \end{theo}
  
The proof  of Theorem \ref{theo:qubit} is provided in Appendix \ref{app:qubit}.   

 With  Theorem  \ref{theo:qubit}   at hand, we can show that maximal activation of the quantum capacity only occurs for channels that are unitarily equivalent to $\map E_{XY}$:  

 \begin{theo}\label{theo:main}  Let  $\map E$ be a  qubit  channel such that $Q  (\map E)  =0$ and    $Q (\map S_\omega  (\map E, \map E))  =1$ for some state $\omega$.   Then,   the channel $\map E$ is unitarily equivalent to  $\map E_{XY}$. 
 \end{theo}

The proof is simple:  by Theorem  \ref{theo:qubit}, we know that channel $\map E$ is either unitary or unitarily equivalent to $\map E_q (\rho)  :=  q  \,  X\rho X +  (1-q)  \,  Y\rho Y$, for some value of $q\in  [0,1]$.  The unitary option is ruled out by the zero-capacity condition $Q  (\map E)  =0$.    Likewise,  the zero-capacity condition rules out  all probability values except $q=1/2$, due to  the hashing bound $Q  (\map E_q)  \ge  1   -    h (q)$ \cite{Devetak05(1)}, where $h(q)  :=  -  q\log q  -  (1-q) \log (1-q)$ is the binary entropy.    Hence, $\map E$ must be unitarily equivalent to $\map E_{1/2}    \equiv  \map E_{XY}$. This concludes the proof of Theorem \ref{theo:main}.

 Theorem \ref{theo:qubit} characterizes all the qubit channels that admit maximal activation of the quantum capacity.  A natural question is whether maximal  activation  can occur for higher dimensional systems.   For a channel with $d$-dimensional input, the maximum value of the capacity is $\log  d$.  Hence, the question is whether  there exists a channel $\map E$ acting on a $d$-dimensional quantum system such that $Q  (\map E)  = 0$ and $Q (\map S_\omega  (\map E, \map E))=  \log d$ for some state $\omega$.     As it turns out, the answer is negative for every $d>2$:    
  
 \begin{theo}\label{theo:nod}  No quantum channel  $\map E$ acting on a $d$-dimensional quantum system with $d>2$ can satisfy the conditions   $Q  (\map E)  =0$ and    $Q (\map S_\omega  (\map E, \map E))  =\log d$ for some state $\omega$. 
 \end{theo}
  The proof of Theorem \ref{theo:nod} is provided in Appendix \ref{app:nod}.

Summarizing, we have shown that switching the order of two uses of a zero capacity channel can yield maximal capacity only for  a specific  type of qubit channels,  that is, channels that are unitarily equivalent to a uniform mixture of the $X$ and $Y$ Pauli gates. For quantum systems of dimension $d>2$,  activation from zero to maximal capacity could still occur through  variants of the quantum SWITCH that permute the order of $N>2$ uses of the given channel \cite{colnaghi2012quantum,facchini2015quantum}.  Finding examples of such activation, however, is  beyond the scope of this paper, which focusses on the $N=2$ case.

   \section{No perfect activation via superposition of independent  noisy channels}\label{path} 
 
\begin{figure}
	\includegraphics[scale=0.88]{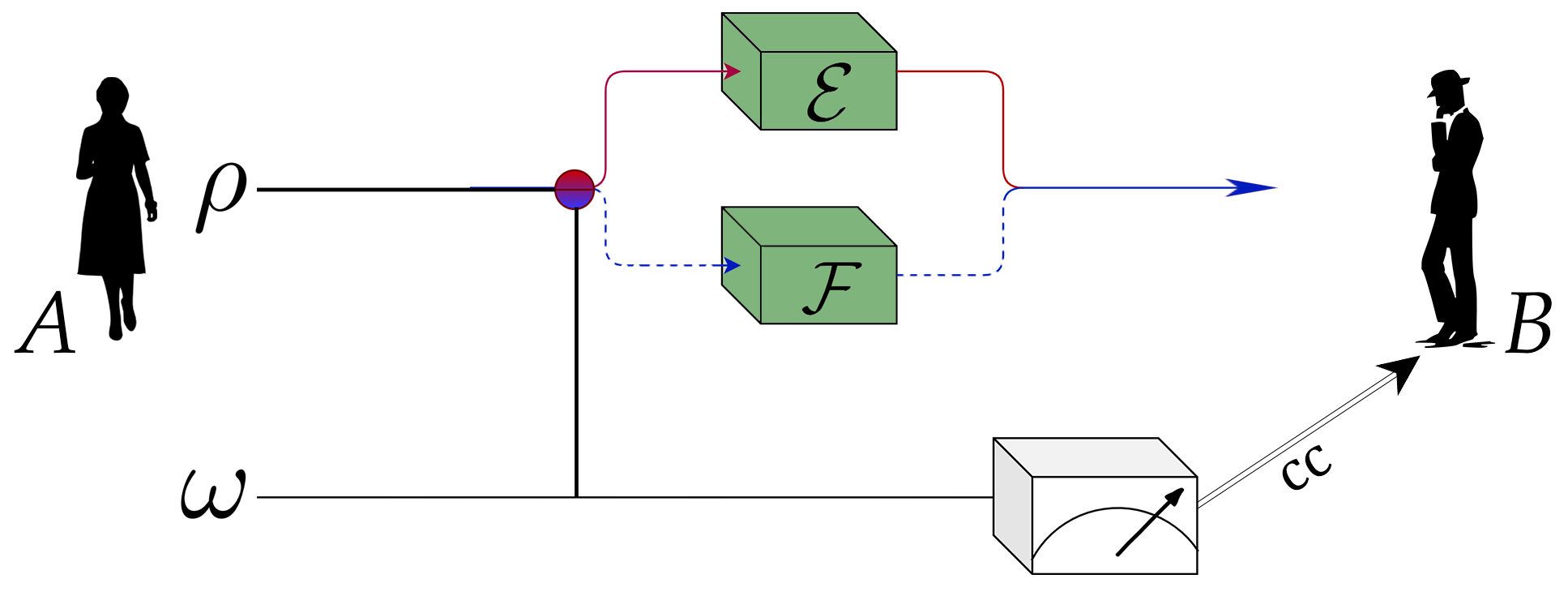}
	\caption{(Color on-line) {\bf Communication through superposition of independent noisy channels.} A communication provider routes a quantum message  on a coherent superposition of two alternative paths between the sender (A) and the receiver (B). The two paths traverse  two independent communication channels $\mathcal{E}$ and $\mathcal{F}$, respectively. The trajectory of the message is controlled by a qubit  (bottom wire in the figure), initialized in the state $\omega$ by the communication provider.   To assist the receiver,  the communication provider measures the  control qubit and communicates the  outcome via a classical transmission line (doubled line in the figure).}
	\label{fig:alternativepaths}
\end{figure}

We now show that the  extreme activation phenomenon shown in the previous section disappears if, instead of combining two independent uses of $\map E_{XY}$ in a superposition of orders, one places them on two alternative paths between the sender and receiver,  sending a message through both paths in a coherent quantum superposition,  as illustrated in Figure \ref{fig:alternativepaths}.    
In other words, if the sender sends a quantum message to the receiver through a superposition of two alternative paths,  and each path leads to an  independent instance of the channel $\map E_{XY}$, then the output state will suffer from  some uneliminable noise. In fact, we prove a much stronger result:  if the sender sends a quantum message  to the receiver through a superposition  of {\em any finite number $N$ of paths}, and if the $N$ paths lead to {\em $N$ independent noisy channels},  then the output state will necessarily suffer from noise. 

The evolution experienced by a quantum particle travelling on a superposition of paths was discussed by Aharonov, Anandan, Popescu, and Vaidman in the unitary case \cite{Aharonov90}. The definition of superposition of quantum evolutions was  subsequently extended to noisy channels in a series of works \cite{Oi03, aaberg2004subspace, Abbott2018, Giulio18}.  
   In the following we briefly review the notion of superposition of noisy channels, following the framework of    \cite{aaberg2004subspace, Giulio18}. 
  
   This framework    is inspired by quantum optics, where a single photon travelling along  $N$ possible paths can be equivalently modelled as the one-photon subspace  of $N$ spatial modes of the electromagnetic field.

 For a generic quantum system $S$ (hereafter called a ``particle''), the superposition of $N$ paths is described by introducing $N$ abstract ``modes". 
  For simplicity,  here we present the framework for $N=2$,  leaving  the extension to arbitrary $N$ to Appendix \ref{app:N}.

 Consider two abstract modes, labelled as $0$ and $1$, each   coming with an internal degree of freedom. In the quantum optics example, the internal degree of freedom is the polarization: each abstract mode is the composite system of  a pair of polarization modes, such as vertical and horizontal polarization,  associated to the same path. 
    A generic quantum  state of mode $m  \in  \{0,1\}$ can be expressed as $|\Psi \>  =  \bigoplus_{n=0}^{n_{\max}}    c_{n}  \,    |\psi_{n}\> $, where $n$ labels the number of particles,  $n_{\max}$ is the maximum number of particles in mode $m$,  $(c_{n})$ are complex amplitudes, and $(|\psi_{n}\>)$ are states of the subspace $\spc H_{n,m}$ associated to $n$ particles in mode $m$. For each mode $m$, we assume that
\begin{enumerate}
\item  the zero-particle subspace $\spc H_{0,m}$ is one-dimensional, meaning that there is a unique vacuum state, hereafter denoted as $|0,m\>$, and 
\item the one-particle subspace  $\spc H_{1,m}$ has dimension $d$,  independently of $m$. 
\end{enumerate}
   
Both assumptions are satisfied in the motivating example of quantum optics, where the vacuum of the electromagnetic field is one-dimensional, and the one-particle subspace associated to each path is a qubit, spanned by the two orthogonal states of horizontal and vertical polarization, respectively.

 Suppose that the evolution  of mode $m  \in  \{0,1\}$  is described by a  quantum channel $\widetilde{\map E}^{(m)}$ that  preserves the number of particles. Preservation of the number of particles implies that the  Kraus operators of $\widetilde{ \map E}^{(m)}$   have the block-diagonal form  $\widetilde E_{i}^{(m)}   =  \bigoplus_{n=0}^{n_{\max}}   \widetilde E_{i,n}^{(m)}$,   where the operator  $\widetilde E_{i,n}^{(m)}$   acts on the $n$-particle subspace $\spc H_{n,m}$ \cite{chiribella2017optimal}. 
  In particular, since the zero-particle subspace is one-dimensional, the operators $\widetilde E_{i,0}^{(m)}$ are complex numbers, called the {\em vacuum amplitudes} of channel $\widetilde{\map E}^{(m)}$ \cite{Giulio18}.  In the following, we will use the shorthand notation  $\alpha_i^{(m)}:  = \widetilde E_{i,0}^{(m)}$.

On the one-particle subspace, the channel $\widetilde{\map E}^{(m)}$ acts as a quantum channel $\map E^{(m)}$ with Kraus operators $E^{(m)}_i  :   =  \widetilde E_{i,1}^{(m)}$.    We call channel  $\widetilde{\map E}^{(m)}$  an {\em extension} of  channel  $\map E^{(m)}$.   In the example of a single photon's polarization, the channel  $\map E^{(m)}$ represents the effective evolution of the polarization  degree of freedom of a single photon travelling  on  the $m$-th path. Instead, the extension  $\widetilde{\map E}^{(m)}$ describes the full evolution of the polarization modes  associated to the $m$-th path.

Now, consider a single particle propagating in a coherent superposition of two possible paths.   The state space of the single particle is the one-particle subspace of the corresponding modes.  A generic state in the one-particle subspace is of  the form 
\begin{align}\label{decomppsi}
|\Psi\>     =   c_0\,  |\psi_0\>\otimes |0,1\>   +  c_1  \, |0,0\>  \otimes |\psi_1\> \, ,     
\end{align}
that is, it is a linear combination of product states where one mode is in a one-particle state and  the other mode is in the vacuum.    The one-particle subspace can be equivalently represented as a bipartite system, whose subsystems are   the internal degree of freedom of the particle (denoted by $S$),  and the particle's path (denoted by $C$).  Explicitly,  the one-particle states can be  written as 
\begin{align}
|\Psi\>     = c_0  \,  |\psi_{0}  \>_S \otimes |0\>_{C}   +  c_1  \,  |\psi_{1}  \>_S \otimes |1\>_{C} \, ,     
\end{align}
 where we associated the orthonormal vectors $|0\>_C$ and $|1\>_C$ to the two possible paths, and we introduced   the notation  $|\psi_0\>_S  \otimes |0\>_C  :  = |\psi_0\>  \otimes  |0,1\> $ and $|\psi_1\>_S  \otimes |1\>_C  :  = |0,0\>  \otimes  |\psi_1\>$.

 Assuming that the  modes $0$ and $1$ evolve independently,   the evolution of the single particle    is simply the restriction of the product channel $\widetilde{\map E}^{(0)} \otimes  \widetilde{\map E}^{(1)} $ to the one-particle subspace.   The action of the Kraus operators $\widetilde E^{(0)}_i \otimes \widetilde E^{(1)}_j$ on a generic state $|\Psi\>$ in the one-particle subspace is 
 \begin{align}
 \left(\widetilde E^{(0)}_i \otimes \widetilde E^{(1)}_j\right)   \, |\Psi\>   & =  c_0  \, E^{(0)}_i  |\psi_0\>   \otimes    \alpha^{(1)}_j  \,|0,1\>   +  c_1  \,  \alpha^{(0)}_i \, |0,0,\> \otimes E^{(1)}_j |\psi_1\>  \, ,   
 \end{align} 
 having used the decomposition (\ref{decomppsi}).   Note that one has  $ \left(\widetilde E^{(0)}_i \otimes \widetilde E^{(1)}_j\right)   \, |\Psi\> =  R_{ij}  \,|\Psi\>$, with 
 \begin{align}
\label{superposition20}
R_{i j } 
  :=      E_i^{(0)}  \otimes        \alpha^{(1)}_{j} \,   |0,1\>\<0,1|  +   \alpha_i^{(0)}  \, |0,0\>\<0,0| \otimes  E_j^{(1)}   \, .            
\end{align}
Hence, the restriction of the product channel  $\widetilde{\map E}^{(0)} \otimes  \widetilde{\map E}^{(1)} $ to the one-particle subspace is the  channel   $\map R  (\widetilde {\map E}^{(0)},  \widetilde {\map E}^{(1)})$ defined by $\map R  (\widetilde {\map E}^{(0)},  \widetilde {\map E}^{(1)})  \,(\rho)  =  \sum_{i,j}  \,  R_{ij} \,\rho  \, R_{ij}^\dag$.  

Regarding the one-particle subspace as the composite system $SC$, made of the internal degree of freedom and the path,  the Kraus operator of the channel  $\map R  (\widetilde {\map E}^{(0)},  \widetilde {\map E}^{(1)})$   can be expressed as 
\begin{align}\label{superposition2}
R_{ij}  =  
     E_i^{(0)}  \,  \alpha_j^{(1)}   \otimes |0\>\<0|_C   +   \alpha_j^{(0)} \,  E_j^{(1)}  \otimes  |1\>\<1|_C\, .          
\end{align}

In the following, we make the standard assumption that the path is initialized in a fixed state $\omega$, independent of the state of the internal degree of freedom $S$ \cite{Gisin05, Oi03, aaberg2004subspace, Abbott2018, Giulio18, Hler20}. Then, the communication between the sender and the receiver is described by  the effective channel   $\map R_\omega   ( \widetilde{\map E}^{(0)} ,\widetilde{\map E}^{(1)})$ defined by the relation
\begin{align}
\map R_\omega   ( \widetilde{\map E}^{(0)} , \widetilde{\map E}^{(1)})  \,   (\rho) :  = \sum_{i,j} \,  R_{ij} \,(\rho \otimes \omega)\,  R_{ij}^\dag \, . 
\end{align}   
We call the channel  $\map R_\omega   ( \widetilde{\map E}^{(0)} , \widetilde{\map E}^{(1)})$   a {\em  superposition   of the channels $\map E^{(0)}$ and $\map E^{(1)}$},  or simply, the {\em superposition channel}.      Note that the superposition  channel  depends not  only on the original channels $\map E^{(0)}$ and $\map E^{(1)}$, but also on the vacuum amplitudes.  Physically, this dependence is due to the fact that the full description of the transmission lines is provided by the channels  $\widetilde{\map E}^{(0)}$ and $\widetilde{\map E}^{(1)}$ acting on the two modes, rather than the channels $\map E^{(0)}$ and $\map E^{(1)}$ acting on the one-particle subspaces of such modes. 
  
   We now show that, if the  one-particle channels $\map E^{(0)}$ and $\map E^{(1)}$ are noisy, then  the superposition channel    $\map R_\omega   ( \widetilde{\map E}^{(0)} , \widetilde{\map E}^{(1)})$ cannot be perfectly corrected.  Hence, any message sent through it will suffer from some unavoidable  noise.   
 This result holds for every finite number of paths:

 \begin{theo} \label{theo:path} Suppose that the state of  a finite-dimensional quantum system is encoded in the internal degree of freedom of a single-particle, which is transmitted through a superposition of $N<\infty$ paths traversing $N$ independent  channels.    If all channels are noisy, then the initial quantum state cannot be retrieved without errors. 
    \end{theo}
The proof is provided in Appendix \ref{app:path}.

An immediate corollary of Theorem \ref{theo:path} is that, for every finite number $N$ and for every given  list of $N$ independent noisy channels $(\map E^{(0)}, \map E^{(1)}, \dots, \map E^{(N-1)})$, the minimum distance between an arbitrary  superposition of these channels and any correctable channel  is nonzero.  Specifically, let us denote by $\map R_\omega   ( \widetilde{\map E}^{(0)} , \widetilde{\map E}^{(1)},  \dots, \widetilde {\map E}^{(N-1)})$ a generic superposition of the channels $(\map E^{(0)}, \map E^{(1)}, \dots, \map E^{(N-1)})$, with generic state $\omega$ and generic extensions $( \widetilde{\map E}^{(0)} , \widetilde{\map E}^{(1)},  \dots, \widetilde {\map E}^{(N-1)})$. Let  $\delta  (\map R_\omega   ( \widetilde{\map E}^{(0)} , \widetilde{\map E}^{(1)},  \dots, \widetilde {\map E}^{(N-1)}))$ be the minimum distance between  channel $\map R_\omega   ( \widetilde{\map E}^{(0)} , \widetilde{\map E}^{(1)},  \dots, \widetilde {\map E}^{(N-1)})$ and the set of correctable channels.     Then, we have the following 
\begin{cor}\label{cor:finitedistance}
For every given list of independent noisy channels $(\map E^{(0)}, \map E^{(1)}, \dots, \map E^{(N-1)})$, there exists a positive number $\delta_*(\map E^{(0)}, \map E^{(1)}, \dots, \map E^{(N-1)})>0$ such that    $\delta  (\map R_\omega   ( \widetilde{\map E}^{(0)} , \widetilde{\map E}^{(1)},  \dots, \widetilde {\map E}^{(N-1)}))   \ge \delta_*(\map E^{(0)}, \map E^{(1)}, \dots, \map E^{(N-1)})$ for every state $\omega$ and for every possible extension $( \widetilde{\map E}^{(0)} , \widetilde{\map E}^{(1)},  \dots, \widetilde {\map E}^{(N-1)})$. 
 In particular, there is a finite non-zero  distance between any superposition of two independent uses of the channel $\map E_{XY}$ and the set of correctable quantum channels. 
\end{cor} 
The proof of Corollary \ref{cor:finitedistance} is provided in Appendix \ref{app:finitedistance}.

Theorem \ref{theo:path} and Corollary \ref{cor:finitedistance} establish a fundamental  result valid for  arbitrary noisy channels and for arbitrary finite numbers of independent uses. Combined with our extreme example of activation, these results strengthen an earlier observation made in \cite{Salek18},  which showed  that the superposition of orders can give rise to  a noiseless {\em heralded}   transmission of quantum states through two entanglement-breaking channels, while the superposition of these two channels cannot.  The key difference is that the perfect quantum communication exhibited by our example takes place {\em deterministically}, and therefore it guarantees a reliable transmission of entanglement over many uses of the channel. In contrast,     heralded  quantum communication can only be used to transmit quantum states involving entanglement among a few particles, because the probability of successful transmission  decays exponentially with the number of particles.   
  
It is worth stressing that  Theorem  \ref{theo:path}  refers to the scenario where the number of paths is finite. When the number of paths $N$ tends to infinity, it is known that nearly perfect quantum communication can sometimes be achieved also  through the superposition of  $N$ independent noisy channels, with an error vanishing as $1/N$  \cite{Giulio18}.    The crucial point, however, is that the quantum SWITCH can achieve perfect deterministic communication with  $N=2$.

\section{Implications of Theorem  \ref{theo:path}}\label{implications}
Theorem  \ref{theo:path} helps understanding the nature of the superposition of orders, by contrasting it with other types of superposition.  

 First of all, the quantum SWITCH of two channels $\map E$ and $\map F$ is {\em not}  a superposition of the channels $\map E^{(0)}  :  = \map E \circ \map F$ and $\map E^{(1)}  :  =\map F\circ  \map E$ regarded as two {\em independent} channels.  
 If it were, then the extreme activation phenomenon shown earlier in this paper would be in contradiction with Theorem \ref{theo:path}. 
  
  Mathematically, the difference between the quantum SWITCH and the superposition of independent channels is evident  from the Kraus representation.   For two independent channels $\map E^{(0)}   =\map E \circ\map F$ and $\map E^{(1)}  =\map F \circ \map E$, the Kraus operators are $E^{(0)}_{ij}  : = E_i  F_j$ and $E^{(1)}_{kl}: = F_k  E_l$, respectively. 
   The superposition of the channels $\map E^{(0)}$ and $\map E^{(1)}$   results into a new channel with  Kraus operators given by Eq. (\ref{superposition}), which now reads  
  \begin{align}
  R_{i,j,k,l} =  E_i F_j  \,  \alpha^{(1)}_{kl}   \otimes |0\>\<0|_C  +     \alpha_{ij}^{(0)} \,  F_k E_l    \otimes |1\>\<1|_C \,,
  \end{align}
   where $(\alpha_{ij}^{(0)})$ and $(\alpha_{kl}^{(1)})$ are  vacuum amplitudes associated to channels $\map E^{(0)}$ and $\map E^{(1)} $, respectively.    The above Kraus operators are  clearly different from the Kraus operators of the channel produced by the quantum SWITCH, shown in   Eq. (\ref{eq1}).

The channel produced by the quantum SWITCH can still be regarded as a ``superposition of the channels $\map E\circ \map F$ and $\map F\circ \map E$'',  in a more general sense discussed in  \cite{aaberg2004subspace, Giulio18}.  This generalized  kind  of superposition is realized by sending a particle on two paths, with   the channel on one path correlated with the channel on the other path.    An explicit realization of the switched channel $\map S  ( \map E, \map F)$ using correlated channels  on two paths has been provided in \cite{Giulio18, kristjansson2020single}  (see also \cite{Oreshkov2018}).   Physically, the correlations between the channels on the two paths can be understood by modelling the quantum channels $\map E$ and $\map F$  as ``collisions'' between the system and two other particles  \cite{Ziman05}, with  the order of the collisions be  determined by  a control qubit.   In this way, the occurrence of a collision realizing channel $\map E$ on one path is anti-correlated with the occurrence of a collision realizing channel $\map E$ on the other path, and similarly for channel $\map F$.  
  From this physical perspective, our result highlights the value of the correlations between the channels on  the two paths as a communication resource.

Theorem  \ref{theo:path} also enables an interesting comparison between the superposition of channel configurations in space and the superposition of channel configurations in time.     Suppose that a communication provider is given two communication devices, which can take as input either one particle or the vacuum.     Let $\widetilde {\map E}$ and $\widetilde {\map F}$ be the two quantum channels describing the two devices. 
One way to use the  devices is to place them in two spatially separated regions, $R_0$ and $R_1$: the communication provider could place channel $\widetilde {\map E}$ in region $R_0$ and channel $\widetilde {\map F}$ in region $R_1$, or the other way round.  By letting the placement of the devices be controlled by a quantum system, the provider could also create a coherent superposition of these two alternative configurations, obtaining a new channel  $\map T (\widetilde {\map E}  ,  \widetilde {\map F}) $ with Kraus operators 
\begin{align}\label{tij}
T_{ij}   =   \widetilde E_i \otimes \widetilde F_j \otimes |0\>\<0|_D   +    \widetilde F_j \otimes \widetilde E_i \otimes |1\>\<1|_D \,,
\end{align} 
where $\{  |0\>_D ,|1\>_D\}$ are orthonormal states of a suitable control qubit $D$.   A single particle could then be sent  in a superposition of two paths, passing through regions $R_0$ and $R_1$, respectively.  Let $|\psi\>_S$ be the initial state of the particle's internal degree of freedom,  $ c_0  |0\>_C +  c_1  \, |1\>_C$ be the initial state of the path, and $d_0  \,  |0\>_D  + d_1\, |1\>_D$    be the initial  state of the qubit controlling the  channels; configuration.   In terms of modes, the state of the particle can be expressed as $c_0\,  |\psi\> \otimes |0,1\>  +  c_1 \,  |0,0\> \otimes |\psi\>$, using the notation of the previous section. The action of the Kraus operator $T_{ij}$ in Eq.~(\ref{tij}) then yields the state
\begin{align}\label{state00}
c_0 d_0   \,  E_i \beta_j  |\psi\>  \otimes |0,1\>\otimes |0\>_D     +   c_0 d_1  \,  \alpha_i  \, F_j  |\psi\>  \otimes |0,1\> \otimes |1\>_D  +  c_1 d_0 \,  |0,0\>  \otimes \alpha_i F_j  |\psi\>  \otimes |0\>_D  +  c_1d_1  \,  |0,0\>  \otimes E_i \beta_j |\psi\>  \otimes |1\>_D  \, ,   
\end{align}
where $\alpha_i$ and $\beta_j$ are the vacuum amplitudes of channels $\widetilde {\map E}$ and $\widetilde {\map F}$, respectively,  $E_i$ and $F_j$ are the Kraus operators of the one-particle restrictions of channels $\widetilde {\map E}$ and $\widetilde {\map F}$, denoted by $\map E$ and $\map F$ respectively. 
The state (\ref{state00}) can be equivalently written as 
\begin{align}\label{state01}
E_i \beta_j  |\psi\>_S  \otimes  |\Phi_0\>_{CD}       +   \alpha_i F_j  |\psi\>_S  \otimes |\Phi_1\>_{CD} \, , 
\end{align}
with 
\begin{align}
|\Phi_0\>  :  =    c_0d_0  \,  |0\>_C  \otimes |0\>_D   +   c_1d_1  \,  |1\>_C  \otimes |1\>_D \qquad {\rm and} \qquad
|\Phi_1\>  :  =    c_0d_1  \,  |0\>_C  \otimes |1\>_D   +   c_1d_0  \,  |1\>_C  \otimes |0\>_D \, .
\end{align}
The state (\ref{state01}) is formally identical to  the state one would get by  sending the particle on two paths,  leading to channels $\map E$ and $\map F$, and associated to the orthogonal states $|\Phi_0\>$ and $|\Phi_1\>$ of a composite control system $CD$.  
In other words, the effective channel acting on the particle's internal degree of freedom   is a superposition of the channels $\map E$ and $\map F$.    By Theorem \ref{theo:path},  no choice of the extensions  $\widetilde {\map E}$ and $\widetilde {\map F}$ can enable a perfect transmission of quantum messages when each of the channels $\map E$ and $\map F$ is noisy.

In contrast, suppose that the regions  $R_0$ and $R_1$ are causally connected, {\em i.e.}  that it is possible to send  signals  from  $R_0$ to  $R_1$.  In particular, this implies that region $R_0$ precedes region $R_1$ in time.  Also, suppose that there exists a mechanism that can place the available communication devices    into regions $R_0$ and $R_1$, so that  the choice of which device  is placed in which region is controlled coherently by a qubit. Such a mechanism could be used to realize  the quantum SWITCH of channels $\widetilde {\map E}$ and $\widetilde {\map F}$.  When a single particle is transmitted,  the quantum SWITCH of channels $\widetilde {\map E}$ and $\widetilde {\map F}$ reduces  to the quantum SWITCH of channels $\map E$ and $\map F$. Hence,  the example provided earlier in the paper shows that perfect quantum communication is possible even if the both channels $\map E$ and $\map F$ are noisy,  

In summary, Theorem \ref{theo:path} can be used to highlight a difference between the coherent placement of two quantum channels on two spatially separated regions, and the  coherent placement of two channels on two causally connected regions. When a single particle is sent, one placement permits perfect quantum communication, while the other does not. Informally, this can be viewed as a difference between superpositions of channel placements in space and superpositions of channel placements in time.

\section{Conclusions}\label{con}

In this work we showed that  the possibility of indefinite causal order  gives rise to an extreme activation phenomenon:  two uses of a zero-capacity quantum channel can be deterministically converted into a single use of a quantum channel with maximal capacity.    Remarkably, such  extreme  form of activation cannot be achieved by sending a particle on a superposition of paths between the sender and the receiver, as long as the processes encountered along different paths are independent and the number of paths is finite.    

Our results are  particularly relevant  in light of the observation that some of the benefits of the superposition of causal  orders  can be obtained also through the superposition of paths in space \cite{Abbott2018}. While the advantages in both scenarios exhibit similarities,   our findings  highlight a fundamental difference between the type of advantages arising from independent channels placed on  a superposition of alternative paths and independent channels placed in a superposition of alternative orders.

\begin{acknowledgments}
This work is supported by the National Natural Science Foundation of China through grant~11675136, the Croucher Foundation, the Canadian Institute for Advanced Research~(CIFAR), the Hong Research Grant Council through grant~17307719 and though the Senior Research Fellowship Scheme SRFS2021-7S02,  the Foundational Questions Institute through grant~FQXi-RFP3-1325,  the John Templeton Foundation through grant  61466, The Quantum Information Structure of Spacetime  (qiss.fr), and the HKU Seed Funding for Basic Research. MB acknowledges support through the research grant of INSPIRE Faculty fellowship from the Department of Science and Technology, Government of India. MA acknowledges support from the CSIR project 09/093(0170)/2016- EMR-I. Research at the Perimeter Institute is supported by the Government of Canada through the Department of Innovation, Science and Economic Development Canada and by the Province of Ontario through the Ministry of Research, Innovation and Science. The opinions expressed in this publication are those of the authors and do not necessarily reflect the views of the John Templeton Foundation.   GC acknowledges stimulating discussions with S Popescu, B Schumacher, P Skrzypczyk, R Renner, J Oppenheim, and V Giovannetti. 

\end{acknowledgments}

\appendix

\section{Proof of Theorem \ref{theo:qubit}}\label{app:qubit}

The proof of Theorem \ref{theo:qubit} is based on three lemmas, whose proofs are provided in the subsequent appendices.

 \begin{lemma}\label{lem:maximalcapacity}
	Let $\map C$ be a generic channel with input $A$ and output $B$, of dimensions $d_A$ and $d_B$, respectively.  Then, the condition $Q(\map C) =  \log d_A$ holds if and only if $\map C$ is correctable, {\em i.e.} if and only if there exists a  correction channel $\map C'$, with input $B$ and output $A$, such that $\map C' \circ \map C  = \map I_A$.      
\end{lemma}

The proof provided in Appendix \ref{a1}. 

Lemma \ref{lem:maximalcapacity} implies that the quantum capacity $Q  (\map S_\omega  (\map E, \map E))$ is maximal if and only if the channel  $\map S_{\omega}   ( \map E, \map E)$ is correctable.     
A necessary condition for the correctability of $\map S_{\omega}   ( \map E, \map E)$ is provided by the following Lemma:
\begin{lemma}\label{lem:pure}  
	Let $\map E$ be a channel from a generic quantum system $A$  (of dimension $d_A  \ge 2$) to itself.
	 If the channel  $\map S_{\omega}   ( \map E, \map E)$  is correctable for some state $\omega$,  then  the channel $\map S_{|\gamma\>\<\gamma|}  (\map E,\map E)$ is correctable for every $|\gamma\>$ in the support of $\omega$, and the same correction channel works for both $\map S_{\omega}   ( \map E, \map E)$ and $\map S_{|\gamma\>\<\gamma|}  (\map E,\map E)$.   
\end{lemma} 

The proof is elementary, and is provided in Appendix \ref{a2} for completeness.

Thanks to Lemma \ref{lem:pure}, we can restrict our attention to the case where the state $\omega$  is pure without loss of generality.  For a pure state $\omega  =  |\gamma\>\<\gamma|$ with  $|\gamma\>  =  c_0  \,  |0\>  +  c_1  \, |1\>$,  the Kraus operators of the channel  $\map S_{\omega}   ( \map E, \map E)$ are 
\begin{align}\label{Kswitch}
S_{ij} =  c_0\,   E_i  E_j  \otimes  |0\>   +  c_1\,  E_j  E_i  \otimes |1\> \,   \, ,
\end{align}
where we used the notation $O  \otimes |\psi\>$ to denote the operator defined by $\big(\, O \otimes |\psi\>\, \big)  \, |\phi\>   :  =   (O|\phi\>)\,     \otimes |\psi\>$, for generic vectors $|\phi\>$ and $|\psi\>$, and for a generic operator $O$.  

Correctability is determined by  the Knill-Laflamme condition \cite{Knill96}, which  reads  
\begin{align}\label{KL}
p  \,  (E_i E_j)^\dag  (E_m  E_n)    +  (1-p)   (E_j E_i)^\dag (E_n E_m)   \,    =  \tau_{mn, ij}  \,  I_A \, ,
\end{align}    
where $\tau$ is a density matrix and $p=  |c_0|^2$.

Now, let us restrict our attention to the qubit case $d_A  = 2$.  In this case, the  Knill-Laflamme condition   implies that the Kraus representation of $\map E$
contains at most two linearly independent operators. 
\begin{lemma}\label{lem:twokraus}
	For every unit vector  $|\gamma\>\in\C^2 $, if the  channel $\map S_{|\gamma\>\<\gamma|}   ( \map E, \map E)$ satisfies the Knill-Laflamme  condition (\ref{KL}), then $\map E$ has at most two linearly independent Kraus operators. 
\end{lemma}
The proof is provided in Appendix \ref{a3}. 
 \medskip 
 
Equipped with the above lemmas, we are now ready to prove Theorem \ref{theo:qubit}.

 \smallskip
 
 {\bf Proof  of Theorem \ref{theo:qubit}.}     Let us start from the``if'' part.     If $\map E$ is unitary, then the channel $\map S_{\omega}   ( \map E, \map E)$ is equal to $\map E^2\otimes \omega$ and can be corrected by discarding the control system and applying the inverse of $\map E$.   Now, suppose that the channel $\map E$ is of the form  $\map E   (\rho)  =  q  \,  (UXU^\dag)\,  \rho\,  (U  X  U^\dag)   +  (1-q)  \,   (U  Y  U^\dag ) \, \rho \,    (U   Y  U^\dag)$,  for some unitary matrix $U$.  
 The switched channel  $\map S_{|\gamma\>\<\gamma|}(\map E, \map E)$  has Kraus operators given by Eq. (\ref{Kswitch}), which reads
 \begin{widetext} 
 	\begin{align}
 	\nonumber 
 	S_{11}    &=  q  \,      I  \otimes  |\gamma_+\>    \qquad  \qquad      &S_{12} & =i  \sqrt{q (1-q)}    \,   U  Z  U^\dag     \otimes |\gamma_-\>            \\     
 	\label{switched} S_{22}    &= (1- q)  \,      I  \otimes  |\gamma_+\>    \qquad  \qquad  &  S_{21} & = -i   \sqrt{q (1-q)}    \,   U Z  U^\dag     \otimes |\gamma_-\>             \, ,
 	\end{align}
 \end{widetext} 
 with $|\gamma_\pm\>    : = c_0  \,  |0\> \pm  c_1\,  |1\>$. 
  
  \medskip

 We now prove the ``only if'' part.  Assume that  there exists a state $\omega$ such that the quantum capacity of the switched channel $\map S_\omega (\map E, \map E)$     is maximal.  Then, Lemma \ref{lem:maximalcapacity} implies that the channel  $\map S_\omega (\map E, \map E)$ is correctable.     Furthermore,   Lemma \ref{lem:pure} implies that the channel                 $\map S_{|\gamma\>\<\gamma|} (\map E, \map E)$ is correctable for every $|\gamma\>$ in the support of $\omega$.      In the following, we will fix one such  state $|\gamma\>$  and we will consider the channel     $\map S_{|\gamma\>\<\gamma|} (\map E, \map E)$.  
 
 Lemma \ref{lem:twokraus} guarantees that channel $\map E$ has a Kraus representation with only two Kraus operators $E_1$ and $E_2$.   Setting $i=j=m=n$ in the Knill-Laflamme condition  (\ref{KL}),  we obtain the relation $(E_i^\dag )^2 (E_i)^2  =  \tau_{iiii} \,  I$, meaning that each operator  $E_i^2$ is proportional to a unitary gate.  
 
 We now  characterize the  operators $O$ such that $O^2$ is unitary. 
 The condition  
 \begin{align}
 (O^\dag)^2  O^2  = I   
 \end{align}
 implies that $O$ is invertible and that one has 
 \begin{align}
 O^\dag O   = (O^\dag )^{-1}   O^{-1}   \, .  
 \end{align}
 In terms of the singular value decomposition $O  = \sum_k  \sqrt{\lambda_k}   \, |v_k\>\<w_k|$, the above relation reads  
 \begin{align}
 \sum_{k}  \lambda_k  \,  | w_k\>\<w_k|    =    \sum_k  \frac1{\lambda_k} \,  |v_k\>\<v_k|  \, .
 \end{align}
 For two-dimensional systems, this means that there are only two possibilities:  
 \begin{enumerate}
 	\item  $\lambda_k=1 \,  \forall k $. In this case,  $O$ is unitary. 
 	\item $\lambda_1 \not = 1$ and $\lambda_2  =  1/ \lambda_1$. In this case, one must have $|v_1\> \propto  |w_2\>$ and $|v_2\> \propto |w_1\>$.  In short,  $O$ is of the form $O =  a  |v_1\>\< v_2|  +  b  |v_2\>\<v_1|$, with $|a|  |b|  =1$.  
 \end{enumerate}

 {\em Case 1.}    If one of the two Kraus operators $E_1$ and $E_2$ is proportional to a unitary matrix, then the normalization condition $E_1^\dag E_1  +  E_2^\dag E_2  =  I$ implies that also the other Kraus operator is proportional to a unitary matrix. Hence, the channel $\map E$ is of the random-unitary form $\map E (\rho)  =   q  \, U_1\rho U_1^\dag  + (1-q)  \,  U_2\rho  U_2^\dag$ for some probability $q\in (0,1)$ and some pair of unitary gates $U_1$ and $U_2$.   Choosing $i=j=1$ and $m=n=2$ in  the Knill-Laflamme condition (\ref{KL}) we obtain  $(U_1^2)^\dag   U_2^2  \propto I$, or equivalently, $U_1^2  \propto  U_2^2$.  Then, there are two possibilities: either $U_1^2\propto   U_2^2  \propto  I$, or the unitaries $U_1$ and $U_2$ have the form $U_1  =   e^{i  \theta_1} \,   \big ( |v_1\>\<v_1|  +   e^{i\theta} \,  |v_2\>\<v_2|  \big) $ and $U_2  =   e^{i  \theta_2} \,  \big ( |v_1\>\<v_1|  -   e^{i\theta}   \,  |v_2\>\<v_2|\big)  $, for some phases $\theta_1,  \theta_2,  \theta \in \R$.     In the second case   the unitaries $U_1$ and $U_2$ commute, and therefore the Knill-Laflamme condition (\ref{KL}) is reduced to the Knill-Laflamme condition for the  channel $\map E^2$. In turn, the correctablity of channel $\map E^2$ implies the correctability of $\map E$, which means that $\map E$ must be unitary, because $\map E$ is a channel from a quantum system to itself.   
 
 The other  possibility is $U_1^2  \propto  U_2^2  \propto I$. This condition means  that the unitaries $U_1$ and $U_2$ are proportional to self-adjoint unitaries, with eigenvalues $+1$ and $-1$.   Since the proportionality constant is an irrelevant global phase, we can discard it without loss of generality.  Hence, we can take  the unitaries $U_1$ and $U_2$ to be self-adjoint. 

 Now,  the Choi operator  of channel $\map E$ is  given by $E   =    \sum_{m,n}  \map E  (|m\>\<n|)  \otimes |m\>\<n|   =   q  |U_1\kk\bb U_1|  +  (1-q)  |U_2\kk\bb U_2| $, using the notation   $|A\kk  =   A_{mn}   \,|m\>\otimes |n\>$, for a generic matrix $A$. 
  Since the unitaries $U_1$ and $U_2$ are self-adjoint, the product $  \bb U_1 |U_2\kk   =  \Tr[U_1 U_2]$ is a real number.    This means that the Gram-Schmidt construction applied to $\{ |U_1\kk,  |U_2\kk\}$ yields an orthonormal basis   $\{  |\Psi_1\>,   |\Psi_2\>\}$ where  the vectors $|\Psi_1\>$ and $|\Psi_2\>$ are linear combinations of $|U_1\kk$ and $|U_2\kk$ with real coefficients. In this basis, the Choi operator can be written as a real symmetric matrix.  Hence, it can be diagonalized as $E  =  \lambda_1  |\Phi_1\>\<\Phi_1|  + \lambda_2  |\Phi_2\>\<\Phi_2|$, where $|\Phi_1\>$ and $|\Phi_2\>$ are  linear combinations of   $|U_1\kk$ and $|U_2\kk$ with real coefficients, and $\<\Phi_1 |\Phi_2\>  = 0$.    Equivalently, the channel $\map E$ can be decomposed as $\map E(\rho)  =   \lambda_1  \,    A_1\rho A_1^\dag  +  \lambda_2   A_2\rho  A_2^\dag$ where $A_1$
 and $A_2$ are real linear combinations of $U_1$ and $U_2$ and $\Tr[A_1  A_2]  =  0$.  
 
 We observe  that every real linear combination of self-adjoint $2\times 2$ unitaries is proportional to a self-adjoint unitary (this is because the  $2\times 2$ self-adjoint unitaries are  of the form   $U  =   \st n \cdot \bs \sigma$ where $\st n \in   \R^3$ is a unit vector, and $\bs \sigma  :  =  (X,Y,Z)$ is the vector with  the three Pauli matrices as entries).   Thanks to this observation, we know that  the operators $A_1$ and $A_2$ are proportional to self-adjoint unitaries, say $A_1  =  \alpha_1  \,  \st n_1\cdot \bs \sigma$ and  $A_2  =  \alpha_2  \,  \st n_2\cdot \bs \sigma$, for proportionality constants $\alpha_i  >0$ and unit vectors $\st n_i \in  \R^3$,  $i \in  \{1,2\}$.       
 Finally, the condition $\Tr [A_1  A_2]= 0$ implies  $\st  n_1\cdot \st n_2  =0$, which in turn implies that the two unitaries $ \st n_1\cdot \bs \sigma$  and $ \st n_2\cdot \bs \sigma$   are of the form $UXU^\dag$ and $UYU^\dag$ for some suitable unitary $U$.

 We now show that the state $\omega$ must be pure.  First of all, we show that every  pure state $|\gamma\>$  in the support of $\omega$ must satisfy the condition $|\<0|\gamma\> |  =  |\<1|  \gamma\>|$.  Indeed, we can set $i=j$ and $m\not = n$ in the Knill-Laflamme condition  (\ref{KL}), obtaining 
 \begin{align}
 p  \, U_1  U_2  +   (1- p)\,   U_2 U_1   \propto I  \, .
 \end{align}
Let us express the two self-adjoint unitaries $U_1$ and $U_2$ as $U_1  =   \st m_1 \cdot \bs \sigma$ and $U_1  =   \st m_2 \cdot \bs \sigma$ for some unit vectors $\st m_1,  \st m_2  \in  \R^3$.    Using the expressions  $ U_1 U_2  =    (\st m_1  \cdot \st m_2)\,  I  +     i   (\st m_1  \times  \st m_2)  \cdot  \bs \sigma$ and  $ U_2 U_1 =    (\st m_1  \cdot \st m_2)\,  I  -     i   (\st m_1  \times  \st m_2)  \cdot  \bs \sigma$ we obtain the condition  
 \begin{align}\label{uno}
 (\st m_1  \cdot \st m_2)\,  I    +    (2p-1)  \,  i   (\st m_1  \times  \st m_2)  \cdot  \bs \sigma  \propto   I  \, .   \,   
 \end{align}
 The second term in the sum is traceless, and therefore orthogonal to the identity operator.   Hence, condition (\ref{uno}) implies  $ (2p-1)  \,  i   (\st m_1  \times  \st m_2)  \cdot  \bs \sigma= 0$.     Since $ \st m_1$ and $\st m_2$ are not  proportional to each other, the only option is to have  $p=1/2$.  Recalling that  $p  =  |c_0|^2$, we obtain that $|c_0|  =  |c_1|  =  \frac 1 {\sqrt 2}$.    
 
 Now, let us show that $\omega$ must be pure. The proof  proceeds  by contradiction. Suppose that $\omega=  \lambda  |\gamma\>\<\gamma|  +  (1-\lambda)  \,  |\gamma'\>\<\gamma'|$, for  two linearly independent states $|\gamma\>$ and $|\gamma'\>$ and for some probability $\lambda  \in  (0,1)$.  Using Eq. (\ref{switched}) for the channels $\map S_{|\eta\>\<\eta|}  (\map E, \map E)$ and $\map S_{|\eta'\>\<\eta'|}  (\map E, \map E)$, we obtain  that the channel  $\map S_\omega  (\map E,\map E)  =  \lambda  \,  \map S_{|\eta\>\<\eta|}  (\map E, \map E) +  (1-\lambda) \, \map S_{|\eta'\>\<\eta'|}  (\map E, \map E)$ has eight Kraus operators
 \begin{widetext} 
 	\begin{align}
 	\nonumber 
 	S_{11}    &=    \sqrt \lambda \,  q  \,      I  \otimes  |\gamma_+\>    \qquad  \qquad      &S_{12} & =  \sqrt{\lambda\, q (1-q)}    \,  \Big [   (\st m_1  \cdot \st m_2)   \,   I   \otimes |\gamma_+\>   + i(\st m_1  \times  \st m_2)  \cdot  \bs \sigma       \otimes |\gamma_-\>         \Big]    \\     
 	\nonumber S_{22}    &= \sqrt \lambda \,   (1- q)  \,      I  \otimes  |\gamma_+\>    \qquad  \qquad  &  S_{21} & =  \sqrt{\lambda \,  q (1-q)}    \,  \Big [   (\st m_1  \cdot \st m_2)   \,   I   \otimes |\gamma_+\>   -  i   (\st m_1  \times  \st m_2)  \cdot  \bs \sigma       \otimes |\gamma_-\>         \Big]      \\
 	\nonumber S_{33}    &=    \sqrt{1-\lambda} q  \,      I  \otimes  |\gamma'_+\>    \qquad  \qquad      &S_{34} & =  \sqrt{(1-\lambda)\, q (1-q)}    \,  \Big [   (\st m_1  \cdot \st m_2)   \,   I   \otimes |\gamma'_+\>   + i(\st m_1  \times  \st m_2)  \cdot  \bs \sigma       \otimes |\gamma'_-\>         \Big]    \\     
 	S_{44}    &= \sqrt {1- \lambda }\,   (1- q)  \,      I  \otimes  |\gamma'_+\>    \qquad  \qquad  &  S_{43} & =  \sqrt{(1-\lambda) \, q (1-q)}    \,  \Big [   (\st m_1  \cdot \st m_2)   \,   I   \otimes |\gamma'_+\>   -  i   (\st m_1  \times  \st m_2)  \cdot  \bs \sigma       \otimes |\gamma'_-\>         \Big]      \, ,
 	\end{align}
 \end{widetext}
 with $|\gamma_+\>  =  |\gamma\>  \, ,  |\gamma'_+\>  =  |\gamma'\>  \, ,   |\gamma_-\>   =   Z |\gamma\>,$ and $|\gamma_-'\>  =   Z  |\gamma'\>$.    Now, the three operators $S_{11},  S_{12}$, and   $S_{33}$ are linearly independent. By Lemma \ref{lem:twokraus}, this implies that the channel $\map S_\omega  (\map E,\map E)$ is not correctable.

 Summarizing, the unit capacity condition $Q (\map S_\omega)  (\map E,\map E)  )=1$ implies one of the following conditions 
 \begin{enumerate}
 \item $\map E$ is unitary, or
 \item  the state $\omega$ is pure and $\map E$ is of the form $ \map E_q   (\rho)  =  q\,  (U  XU^\dag) \rho (U X U^\dag)  +  (1-q)\,  (U  YU^\dag) \rho (U Y U^\dag)$ for some suitable unitary matrix $U$. 
  \end{enumerate}
 
 \medskip 
 
 {\em Case 2.}  The channel $\map E$ is of the form $\map E (\rho)   =  A\rho A^\dag +  B\rho B^\dag$, with $A   =    a  |v_1\>\<v_2|  +  b |v_2\>\<v_1| $ and $B   =    c  |v_1\>\<v_2|  +  d |v_2\>\<v_1| $ and $|a|^2  + |c|^2   =  |b^2|+  |d^2|  =1$.    Its Choi operator is $E   =  |A\kk\bb A|  +  |B\kk \bb B|$, where the vector $|A\kk  \in  \C^2\otimes \C^2$ is defined as  $|A\kk  =   (A\otimes I)  \,  |I\kk$, with $|I\kk  = \sum_n  |n\>\otimes |n\>$.  In the two-dimensional subspace spanned by the vectors $|v_1\>  \otimes |\overline v_2\>$ and $|v_2\>\otimes |\overline v_1\>$, the Choi operator has the matrix representation 
 \begin{align}
 \nonumber E  &=  \begin{pmatrix}
 |a|^2  &   \overline a  b  \\
 \overline b a &  |b|^2    
 \end{pmatrix}  + 
 \begin{pmatrix}
 |c|^2  &   \overline c  d  \\
 \overline d c &  |d|^2    
 \end{pmatrix}  \\
 \nonumber &  =   \begin{pmatrix}
 1&   c  \\
 \overline c &  1 
 \end{pmatrix}
 \qquad c:  =   \overline a b  + \overline c d  \\
 \label{CC'} &  =   \begin{pmatrix}
 1&   0  \\
 0 &  e^{-i\theta} 
 \end{pmatrix}
 E'
 \begin{pmatrix}
 1&   0  \\
 0 &  e^{i\theta}  
 \end{pmatrix}
 \, , \quad e^{i\theta}   =  \frac c{|c|} \, , \quad   E'  = \begin{pmatrix}
 1&   |c|  \\
 |c| &  1 
 \end{pmatrix} \, .
 \end{align} 
 Now, the matrix $E'$ can be expressed  as $E'    =   \frac {1+|c|}{2}  \, \begin{pmatrix}
 1&   1  \\
 1 &  1 
 \end{pmatrix}   +   \frac {1-|c|}{2}   \, \begin{pmatrix}
 \phantom{-}1&   -1  \\
 -1 &  \phantom{-}1 
 \end{pmatrix} $. This means that $E'$ is the Choi operator of the random unitary channel  
 \begin{align}
 \map E'  (\rho)   =  \frac {1+|c|}{2}    \,   U'_1  \rho  U_1^{\prime \dag }  +    \frac {1+|c|}{2}  \,  U'_2\rho  U_2^{\prime \dag}  \, ,
 \end{align} 
 with $U'_1   =   |v_1\>\<v_2|  +   |v_2\>\<v_1|$ and $U'_2   =   |v_1\>\<v_2|  -   |v_2\>\<v_1|$.  From Eq.(\ref{CC'}), one can see that the channel $\map E$ is given by 
 \begin{align}
 \map E  (\rho)   =  \frac {1+|c|}{2}    \,   U_1  \rho  U_1^{\dag }  +    \frac {1-|c|}{2}  \,  U_2\rho  U_2^{ \dag}  \, ,
 \end{align} 
 with $U_1   =   |v_1\>\<\widetilde v_2|  +   |\widetilde v_2\>\<v_1|$,  $U_2   =   |v_1\>\<\widetilde v_2|  -   |\widetilde v_2\>\<v_1|$, and $|\widetilde v_2\>   =  e^{-i\theta} \,  |v_2\>$.     In summary, the channel $\map E$ is random unitary. This brings us back to {\em Case 1}.

 \qed

\section{Proof of Lemma \ref{lem:maximalcapacity}}\label{a1}

\Proof  The ``if" part is trivial: clearly, a correctable channel has capacity $Q (\map C)  =  \log d_A$.    For the ``only if'' part, we use the Holevo-Werner upper bound   $Q(\map C)  \le  \log \|    \map  T_B  \circ \map C \|_{\diamond}$, where $\map T_B$ denotes the transpose map on system $B$, and  $\| \Delta \|_{\diamond}$ denotes  the diamond norm of  a generic Hermitian-preserving map $\Delta$ \cite{Holevo99}.  Note that the upper bound can be equivalently written as $ Q  (\map C)  \le  \log \|  \map D  \circ  \map T_A\|_{\diamond}$ with $\map D  =   \map T_B\circ \map C \circ \map T_A$. 

Now, suppose that $Q (\map C)  =  \log d_A$.  Using the notation   $|A\kk  =   (A\otimes I)  \,  |I\kk$,  $|I\kk  = \sum_n  |n\>\otimes |n\>$, we obtain 
\begin{widetext} 
	\begin{align}
	\nonumber d_A    &  =    2^{Q(\map C)}   \\
	\nonumber &\le    \| \map D  \circ \map T_A\|_\diamond \\
	\nonumber &   =  \sup_{  |\Psi\kk  \in  \spc H_A\otimes \spc H_A}  \,    \big\|   (\map I_A\otimes \map D  \circ \map T_A)  (   |\Psi\kk\bb\Psi|) \, \big\|_1 \\
	\label{ineq1}
	&\le  \sup_{  |\Psi\kk  \in  \spc H_A\otimes \spc H_A}  \,    \big\|   (\map I_A\otimes   \map T_A)  (   |\Psi\kk\bb\Psi|) \, \big\|_1 \\
	\nonumber &   =  \sup_{\Psi  \in  {\sf L} (\spc H_A) \, ,   \Tr[\Psi^\dag\Psi]=1}     \left\|       A^{(\Psi)}_+   -         A^{(\Psi)}_-   \, \right\|_1  \, , \qquad A^{(\Psi)}_{\pm}   =    (  \Psi\otimes I_A) \,   P_{\pm} (\Psi^\dag \otimes I_A) \, , \quad  P_\pm  =  \frac{  I^{\otimes 2}   \pm  {\tt SWAP}}2   \\
	\label{ineq2} &   \le  \sup_{\Psi  \in  {\sf L} (\spc H_A) \, ,   \Tr[\Psi^\dag\Psi]=1}     \left\|       A^{(\Psi)}_+  \right\|_1  +         \left\|  A^{(\Psi)}_- \, \right\|_1\\
	\nonumber &   =     \sup_{\Psi  \in  {\sf L} (\spc H_A) \, ,   \Tr[\Psi^\dag\Psi]=1}     \Tr  [( \Psi^\dag \Psi \otimes  I_A)  P_+]    +   \Tr  [( \Psi^\dag \Psi \otimes  I_A)  P_-]\\
	& = d_A \, ,   \label{ineq3}  
	\end{align}
\end{widetext}
where Equation (\ref{ineq1}) follows from the contractivity of the trace norm under the action of quantum channels and Equation (\ref{ineq2}) follows from the triangle inequality of the trace norm.   

Now, in order for Equation (\ref{ineq3}) to hold,  all intermediate inequalities must be saturated.  Inequality (\ref{ineq2}) is saturated if and only if the operators $A_+^{(\Psi)}$ and $A_-^{(\Psi)}$ have orthogonal support, that is, if and only if  $\Tr\left[ A_+^{(\Psi)}   A_-^{(\Psi)}\right]   =0$.  Explicitly, we have  
\begin{widetext} 
	\begin{align}
	\nonumber \Tr\left[ A_+^{(\Psi)}   A_-^{(\Psi)}\right]     &  =  \Tr[  (  \rho   \otimes I_A) \,   P_+ (\rho  \otimes I_A)   \,   P_- ] \, , \qquad \rho  =  \Psi^\dag \Psi \\
	&  \ge \Big(\< \phi| \otimes \<\phi |  \rho  \Big) \,   P_-   \,  \Big(|\phi\>\otimes \rho |\phi\>\Big)  \, ,  \qquad \forall  |\phi\>  \in \spc H_A \, , \<\phi|\phi\>  = 1 \, .
	\end{align} 
\end{widetext}
Note that the right-hand side is zero if and only if $\rho  |\phi\>$ is proportional to $|\phi\>$ for every $|\phi\>  \in \spc H_A$, that is, if and only if $\rho   =  I_A/d_A$.     In other words, the state $|\Psi\kk$ must be maximally entangled. 

For the canonical maximally entangled state $|\Psi\>   =  \sum_{n=1}^{d_A} \,|n\>\otimes |n\>/\sqrt{d_A}$, the Holevo-Werner bound yields the chain of inequalities
\begin{align}
\nonumber d_A   &  \le   \left \|       ( \map I_A\otimes \map D\circ \map T_A )(  |\Psi\>\<\Psi|)    \right\|_1    \\
\nonumber   &   = \frac 1 {d_A}  \,   \left  \|   ( \map I_A\otimes \map D )  (P_+)    -   ( \map I_A\otimes \map D )  (P_-)   \right\|_1   \\
\label{ineq4}   &   =   \frac 1 {d_A}  \,   \Big  (  \left  \|   ( \map I_A\otimes \map D )  (P_+) \right\|_1   - \left\|  ( \map I_A\otimes \map D )  (P_-)   \right\|_1\Big)   \\ 
\nonumber   &   \le \frac 1 {d_A}  \,     \big (  d_+  +  d_-\big) \\
&=d_A \, ,
\end{align}
which again implies that all inequalities must be saturated.   In particular, the triangle inequality  (\ref{ineq4}) must hold with the equality sign, meaning that the operators  $( \map I_A\otimes \map D )  (P_+)$ and $ ( \map I_A\otimes \map D )  (P_-) $ must have orthogonal support, namely 
\begin{align}\Tr\big[  ( \map I_A\otimes \map D )  (P_+)\,  ( \map I_A\otimes \map D )  (P_-)\big]      =  0 \, .
\end{align}       Expanding the channel $\map D$ in  a Kraus representation $\map D  (\rho)  =  \sum_{i}  D_i \rho D_i^\dag$,  and using the fact that each map $D_i\cdot D_i^\dag$ is completely positive, we obtain the condition 
\begin{align}
\Tr\Big[  (I_A  \otimes   D_j^\dag D_i)  \,  P_+  (I_A  \otimes   D_i^\dag D_j)   P_-\Big]  =  0  \qquad \forall i, \forall j \, ,  
\end{align}
which in turn implies
\begin{align}
\Big (  \< \phi | \otimes  \<  \phi|   D_i^\dag D_j \Big)     \, & P_- \,  \Big(   |\phi\>  \otimes  D_j^\dag D_i |\phi\> \Big)  =  0 \nonumber \\ &\qquad \forall i\, , \forall j   \, , \forall |\phi\>\in\spc H_A \, ,  \<\phi|\phi\>  = 1 \, .
\end{align} 
The above equation is satisfied if and only if the vector $D_j^\dag D_i |\phi\>$ is proportional to $|\phi\>$, that is, if and only if $D_j^\dag D_i  =  \tau_{ij} \,     I$, for some proportionality constant $\tau_{ij}  \in  \C$.  This is nothing but the Knill-Laflamme condition for error correction \cite{Knill96}. Hence, there must exists a correction channel $\map D'$ such that $\map D' \circ \map D =  \map I_A$. Recalling that $\map  D$ is equal to $\map T_B \circ \map C \circ \map T_A$, we then obtain the chain of equalities  
\begin{align}
\nonumber \map I_A   &=     \map T_A  \circ \map T_A  \\
\nonumber    &   =       \map T_A  \circ   ( \map D'  \circ \map D)  \circ \map T_A  \\ 
\nonumber    &   =       (\map T_A  \circ    \map D'  \circ  \map T_B   )  \circ \map C  \circ   (\map T_A  \circ \map T_A)  \\ 
&  =   \map C'  \circ \map C   \, , \qquad \map C'     :=   \map T_A  \circ \map R \circ \map T_B \, .
\end{align}
Since $\map C'$ is a quantum channel, we conclude that $\map C$ is correctable.  \qed  
\medskip

\section{Proof of Lemma \ref{lem:pure}}\label{a2}

\Proof    If  $|\gamma\>$ is in the support of $\omega$,  then $\omega$ can be decomposed as $\omega =    t  \,  |\gamma \>\<\gamma| +  (1-t)   \sigma$, where  $t>0$ is a non-zero probability and $\sigma$ is a suitable density matrix.   By linearity, one has  $\map S_{\omega}   ( \map E, \map E)  =  t\,  \map S_{|\gamma\>\<\gamma|}  (\map E,\map E)  +  (1-t) \,  \map S_{\sigma}   ( \map E, \map E)$.  Now, let $\map C'$ be a correction for $\map S_{\omega}   ( \map E, \map E)$.  The decomposition of $\map S_{\omega}   ( \map E, \map E)$ implies the condition 
\begin{align}
\nonumber  \map I_A    &  =  \map C' \circ \map S_{\omega}   ( \map E, \map E)   \\
&   =  t \, \map C'\circ\map S_{|\gamma\>\<\gamma|}  (\map E,\map E)  +  (1-t)\, \map C' \circ \map S_{\sigma}   ( \map E, \map E) \, .
\end{align} 
Since the identity is an extreme point of the set of quantum channels, the above condition implies $\map C'\circ\map S_{|\gamma\>\<\gamma|}  (\map E,\map E)  =  \map I_A$. This proves that $\map S_{|\gamma\>\<\gamma|}  (\map E,\map E)$ is correctable and admits the same correction  as  $\map S_{\omega}   ( \map E, \map E)$.   \qed

\medskip

\section{Proof of Lemma \ref{lem:twokraus}}\label{a3}

\Proof  For an arbitrary channel $\map C$ with arbitrary input and output Hilbert spaces $\spc H_{\rm in}$ and $\spc H_{\rm out}$,  error correction on arbitrary inputs is possible only if  quantum packing bound  
\begin{align}
d_{\rm out}  \ge r   \,  \times d_{\rm in} \, ,
\end{align}
is satisfied (see {\em e.g.} \cite{Giulio11}), where $d_{\rm out}$ and $d_{\rm in}$ are the dimensions of $\spc H_{\rm out}$ and $\spc H_{\rm in}$, respectively, and $r$ is the number of linearly independent Kraus operators of the channel $\map C$.   For the switched channel $\map S_{|\gamma\>\<\gamma|}   ( \map E, \map E)$, we have $d_{\rm in }=  d$  and $d_{\rm out}  =  2 d$. Hence, we have the bound 
\begin{align}
r_{\rm switch} \le 2 \, , 
\end{align} 
where $r_{\rm switch}$ is the number of linearly independent Kraus operators of $\map S_{|\gamma\>\<\gamma|}   ( \map E, \map E)$.    We now show that also the original channel $\map E$ can have at most $2$ linearly independent Kraus operators.  To this purpose, consider the Knill-Laflamme condition (\ref{KL}) and set   $i=j=m=n$.   With this choice, we obtain $(E_i^\dag)^2   (E_i)^2  =  \tau_{ii,ii} \,  I$ for every $i$, which implies that each non-zero Kraus operator $E_i$ is invertible.     Now, suppose that $\map E$ has $r$ linearly independent Kraus operators  $( E_i)_{i=1}^r$.  For every fixed $j$,  the operators $(E_i  E_j)_{i=1}^r$ must be linearly independent, and so must be the operators $(E_jE_i)_{i=1}^r$.  Hence, also the operators $   (  c_0   \,  E_iE_j  \, \otimes |0\>  +    c_1   \,   E_j  E_i \otimes |1\>)_{i=1}^r$ must be linearly independent.  This means that the switched channel $\map S_\omega  (\map E,  \map E)$  has at least $r$ linearly independent Kraus operators, namely  
\begin{align}
r_{\rm switch}  \ge r\, . 
\end{align}
In conclusion, we obtained the bound $r\le 2$.  \qed

\section{Proof of Theorem \ref{theo:nod}}\label{app:nod}  

\Proof  Let $\map E$  be a generic quantum channel with $d$-dimensional input system $A$ and $d$-dimensional output system $B$, with $d>2$.   The maximal capacity  condition  $Q(\map S_\omega (\map E, \map E) )  =  \log  d$  implies that channel $\map E$ has at most two linearly independent Kraus operators (by Lemmas \ref{lem:maximalcapacity} and \ref{lem:twokraus}).    
Hence, the Choi operator $E  :  =  (\map E \otimes \map I)  (  |I\kk \bb I| )$ has rank at most 2, and channel  $\map E$ has a Kraus representation of the form $\map E  (\rho)   =E_1  \rho E_1^\dag+  E_2 \rho E_2^\dag$ with $E_1^\dag E_1  + E_2^\dag E_2  = I$. 

Now, we show that the channel $\map E$ cannot have zero capacity for any $d>2$.  To this purpose, we use the fact that the quantum capacity is lower bounded by the  the coherent information of the channel, which in turn is lower bounded by the coherent information  the Choi state $ E/d$. In formula, 
\begin{align}\label{lowerboundcapacity}
Q (\map E)  \ge  I_{\rm c}  (\map E)  \ge    I_{\rm c}  (  A  \> B)_{E/d} \, .
\end{align}  
Then, it suffices to show that the Choi state has non-zero coherent information.  Explicitly, the coherent information of the Choi state is $I_{\rm c}  (  A  \> B)_{E/d}       =   S (\rho_B)   -  S  (\rho_{AB})$, with   $\rho_{AB}  =  E/d$ and
\begin{align} 
\nonumber \rho_{B}    &= \Tr_A  [   E/d]   \\
  &   =  \frac{ E_1  E_1^\dag  +  E_2^\dag E_2}{d}  \, .
\end{align}
Since the Choi state has rank at most $2$,  and its von Neumann entropy is at most $\log 2  = 1$, and we have the bound
\begin{align}\label{boundchoi}
 I_{\rm c}  (  A  \> B)_{E/d}     &\ge  S    \left( \rho_B \right)  -  1  \, .
\end{align}

At this point, we recall the normalization condition $E_1^\dag E_1  +  E_2^\dag E_2  =  I$. Defining $P:  =  E_1^\dag E_1$, we then have $E_2^\dag E_2  =  I  - P$. Moreover, we recall that, for every operator $A$, the operators $A^\dag A$ and $A A^\dag$ are unitarily equivalent. Hence, there exist two unitary operators $U_1$ and $U_2$ such that $  E_1E_1^\dag =   U_1  P  U_1^\dag$ and $E_2  E_2^\dag  =  U_2  (  I-  P)   U_2^\dag  =    I  -  U_2  P  U_2^\dag $.  
 The state $\rho_B$ can then be written as 
 \begin{align}\label{rhoB}
 \rho_B   =  \frac  {  I    +   U_1  P  U_1^\dag   -  U_2  P  U_2^\dag }d  \, ,
 \end{align}
and its operator norm $\|  \rho_B\|_\infty$ (equal to its maximum eigenvalue) satisfies the bound 
\begin{align}
\nonumber \|  \rho_B \|_\infty &=    \frac { 1  +  \max_{|\psi\> :  \,  \|\psi\|=1}   \<\psi|        U_1  P  U_1^\dag   -  U_2  P  U_2^\dag  |\psi\>}d  \\
  & \le  \frac  2 d \, .  \label{bound2d}
\end{align}
Using this fact, we can lower bound the min-entropy $S_{\min}  (\rho_B)   :  =  - \log \|  \rho_B\|_\infty$ as  $S_{\min} (\rho_B) \ge \log (d/2)  = \log d   -  1$.  
Since the min-entropy is a lower bound to the von Neumann entropy, we obtain the bounds 
\begin{align}\label{minentropy}
S  (\rho_B)  \ge S_{\min}  (\rho_B)  \ge \log d -1 \,,
\end{align} 
 and 
\begin{align}
 \nonumber I_{\rm c}  (  A  \> B)_{E/d}     &\ge  S    \left( \rho_B \right)  -  1  \\
   &  \ge   \log d - 2  \, . \label{ineq}
\end{align}
For $d>4$, this bound implies that the coherent information of the Choi state is  strictly positive. In this case,  Eq. (\ref{lowerboundcapacity}) implies that the quantum capacity is also strictly positive.    

To conclude the proof, we consider separately the cases of $d=4$ and $d=3$.  For $d=4$, the proof is based on the bound (\ref{ineq}), which guarantees that the coherent information of the Choi state is larger than, or equal to zero.  If the coherent information is larger than zero, then Eq. (\ref{lowerboundcapacity}) implies that the quantum capacity is  strictly positive.   It remains to analyze the case where  the coherent  information is zero, namely the case in which the bound  (\ref{ineq}) is attained with the equality sign.  
To achieve  the equality,  one must  have the equality sign in Eq. (\ref{bound2d}), meaning that the maximum eigenvalue of $\rho_B$ is exactly $2/d$.   Moreover, one must  have the equality in the bound  $S(\rho_B)   \ge S_{\min} (\rho_B)$.   Such equality is attained only when $\rho_B$ is proportional to a projector. Since one of the eigenvalues is $2/d$, we conclude that the projector has rank $r=  d/2  =  2$.    From the definition of $\rho_B$ in  Eq. (\ref{rhoB}), we then obtain that the operator $P$ should be a projector on a two-dimensional subspace.   Finally, recall the definition $P =  E_1^\dag E_1$, which implies $E_1  =   U_1   \sqrt {P}  =  U_1  P$ for some unitary operator $U_1$. For every state $\rho$ with support contained  into the support of $P$, one has $\map E  (\rho)   =  E_1  \rho E_1^\dag   =   U_1 \rho U_1^\dag$. Since the channel acts unitarily on a two-dimensional subspace, its quantum capacity  is at least $1$.   In summary, for $d=4$ any quantum channel $\map E$ satisfying the condition $\map S_\omega (\map E, \map E)$ for some state $\omega$  must have  non-zero quantum capacity.

Let us move now to the $d=3$ case. By Lemma \ref{lem:maximalcapacity}, the maximal capacity condition  $Q  (\map S_\omega (\map E, \map E) )  =  \log d$  implies that the channel $\map S_\omega (\map E, \map E)   $ must be correctable, namely that  there exists a channel $\map C'$ such that $\map C' \circ \map S_\omega (\map E, \map E)   =  \map I$.  From Eq. (\ref{switchsame}) in the main text, we have  $\map S_\omega (\map E, \map E)   (\rho)  =   (    \sum_{i,j}   \{  E_i,  E_j\} \rho  \{  E_i,  E_j\}   \otimes \omega   +      [  E_i,  E_j] \rho  [ E_i,  E_j]   \otimes Z\omega Z \, )/4$ for every density matrix $\rho$.    Hence, there must exist  constants $\lambda_{ij}$ such that  
\begin{align}
  \map C'  \circ  \Big (    \{  E_i,  E_j\}  (\cdot)  \{  E_i,  E_j\}   \otimes \omega \Big)  =  \lambda_{ij}\,    \map I  \, .
\end{align}
  Equivalently, there must exist a channel $\map E'$ such that 
  \begin{align}
    \map E'    \circ \Big (    \{  E_i,  E_j\}  (\cdot)  \{  E_i,  E_j\}\Big)   =  \lambda_{ij}\, \map I \, .  
  \end{align} 
Now, consider the completely positive map $ \{  E_i,  E_j\}  (\cdot)  \{  E_i,  E_j\}$.  Since this map transforms system $S$ into itself, and is correctable, it must be proportional to a unitary channel. Hence, each operator $\{  E_i,  E_j\}$ must be proportional to a unitary.  In particular,  $E_1^2$ and $E_2^2$ must be proportional to unitaries.      

Now,  let us express $  E_1$ and $E_2$ as $E_1  =  U_1  \sqrt{  P_1}$  and $ E_2  = U_2 \sqrt{P_2}$, where $U_1$ and $U_2$ are unitaries, $P_1  :=  E_1^\dag E_1$, and $P_2   :=  E_2^\dag E_2   $.  For $i\in \{1,2\}$, the condition that $E_i^2$ is proportional to a unitary can be written as $E_i^\dag  =  \lambda_i   \,   V_i$, for some constant $\lambda_i$ and some unitary operator $V_i$.   

We now show that, if  the constant $\lambda_i$ is zero for some $i$, then the quantum channel has non-zero capacity.   To see that this is the case, note that the condition $\lambda_i  = 0$ implies $  U_i \sqrt{  P_i} U_i\sqrt {P_i}   =  0$, and also  $\sqrt{  P_i} U_i\sqrt {P_i}   =  0$. The last condition implies that the kernel of $\sqrt {P_i} $ contains all vectors of the form $ U_i  |\psi\>$, where $|\psi\>$ is in the support of $\sqrt {P_i}$.  Hence, the dimension of the kernel of $\sqrt{P_i}$ cannot be smaller than the dimension of the support of $\sqrt{P_i}$.  Since the total dimension is $d=3$, this condition implies that the kernel has dimension $2$ and the support has dimension $1$.  In other words, the operator $P_i$ has the form $P_i   =  |\eta_i\>\,\eta_i|$ for some (possibly subnormalized) vector $|\eta_i\>$.  

Now, recall the definition $P_i   :=  E_i^\dag E_i$ and the normalization condition $P_1  +  P_2  = I$.  If $P_1  =  |\eta_1\>\<\eta_1|$, then   $P_2   =    I  -  |\eta_1\>\<\eta_1|$, and $P_2$ acts as a projector in the two-dimensional subspace orthogonal to $|\eta_1\>$.   For every vector $|\psi\>$ in such subspace,   one has $  \map E  (|\psi\>\<\psi|)   =  U_2   |\psi\>\<\psi| U_2^\dag$.  Hence, the quantum capacity of $\map E$ is at least $1$.  Similarly, if $P_2  =  |\eta_2\>\<\eta_2|$, then   $P_1   =    I  -  |\eta_2\>\<\eta_2|$, and $P_1$ acts as a projector in the two-dimensional subspace orthogonal to $|\eta_2\>$.  For every vector $|\psi\>$ in such subspace,   one has $  \map E  (|\psi\>\<\psi|)   =  U_1   |\psi\>\<\psi| U_1^\dag$.  Hence, the quantum capacity of $\map E$ is at least $1$.  

Summarizing, the quantum capacity is non-zero whenever $\lambda_1=0$ or $\lambda_2 =  0$.  Now, consider the case when  $\lambda_i$ is non-zero for every $i\in \{1,2\}$.  In the following we will show that, also in this case, the capacity of $\map E$ is non-zero.  

First, note that the condition $E_i^2  =  \lambda_i \,  V_i$ with $\lambda_i\not  = 0$ implies that $E_i$ is an invertible matrix. Since $E_i  =  U_i  \sqrt{P_i}$, also $\sqrt{P_i}$ must be an invertible matrix.   Moreover, the condition $E_i^2  =  \lambda_i \,  V_i$ implies  
\begin{align}
\nonumber |\lambda_i|^2  \,   I    &   =  \left(  E_i^{2} \right)^\dag  \,  E_i^2   \\  
&  =   \sqrt {P_i} \, U_i^\dag  \,  P_i\,    U_i  \, \sqrt{P_i}   \, ,     
\end{align}   
or equivalently  
\begin{align}
 U_i^\dag  \,  P_i\,    U_i    =  |\lambda_i|^2  \, P_i^{-1} \, .
\end{align}
The last equation implies that $P_i$ and $|\lambda_i|^2  \, P_i^{-1}$ have the same spectrum. 

Let $(a,b,c)$ be the eigenvalues of $P_1$, listed in descending order $a\ge b\ge c$.    Then, the eigenvalues of $P_1^{-1}$ are $\left(\frac1c, \frac 1b , \frac 1a\right)$, still listed in descending order. The condition that $P_1$ and $|\lambda_1|^2  \, P_1^{-1}$ have the same spectrum implies  $|\lambda_1|^2  =  b^2$ and $c=  b^2/a$.  Summarising, the spectrum of $P_1$  is of the form  $(a,  b, c)$ with $a\ge b\ge c\equiv  b^2/a$.    Similarly, the spectrum  of $P_2$ must be of the form $  (a',  b' ,  c')$ with $a'\ge b'\ge c' \equiv  b^{\prime 2}/a'$.  On the other hand, the condition $P_1  +  P_2  =  1$ implies $a'  =  1-c$,  $b' = 1- b$, and $c'  =  1-a$.   Hence, we must have 
\begin{align}1-a  = c'  =   \frac{b^{\prime \, 2}}{a'}    =    \frac{  (1-b)^2}{1- c}   =  \frac{  (1-b)^2}{  1-  \frac{  b^2} a} \,,
\end{align} that is,    $(a-b)^2   =  0$.       This condition implies $a=  b  =  c$, and $a'=b'=c'$, meaning that the operators   $P_1$ and $P_2$ are proportional to the identity.  Hence, the operators $E_1 =  U_1 \sqrt{P_1}$ and $E_2=  U_2 \sqrt{P_2}$ are proportional to unitaries. 

From Eq. (\ref{rhoB}) and from the definition  $P:  =  E_1^\dag E_1 \equiv P_1$  we obtain the equality $\rho_B   =  I/d$.  Hence, the bound (\ref{boundchoi}) becomes 
\begin{align}
 I_{\rm c}  (  A  \> B)_{E/d}      \ge  \log d   -1  =  \log 3  -  1  >  0  \, .   
\end{align}
Since the coherent information of the Choi operator $E/d$ is a lower bound to the quantum capacity, we proved  that  the channel $\map E$ has non-zero capacity.    
    
Summarizing, any quantum channel $\map E$ acting on a $d$-dimensional quantum system with $d> 2$ and satisfying the condition $Q(\map S_\omega (\map E, \map E)) =  \log d$ for some state $\omega$ must have $Q  (\map E)  >0 $. This concludes the proof of the theorem.  \qed  

\section{Superpositions of  $N$ independent channels}\label{app:N}

Here we review the notion of superposition of $N$ independent channels, following the  framework of \cite{aaberg2004subspace, Giulio18}. In this framework, the superposition of $N$ paths is described by introducing $N$ abstract ``modes", hereafter labelled by  an index $m$ taking values between $0$ and $N-1$. 
     
Each abstract mode  comes with an internal degree of freedom, such as the photon degree of freedom for spatial modes in quantum optics.  A generic quantum  state of mode $m$ can be expressed as $|\Psi \>  =  \bigoplus_{n=0}^{n_{\max,m}}    c_{n}  \,    |\psi_{n}\> $, where $n$ labels the number of particles,  $n_{\max, m}$ is the maximum number of particles in mode $m$,  $(c_{n})$ are complex amplitudes, and $(|\psi_{n}\>)$ are states of the subspace $\spc H_{n,m}$ associated to $n$ particles in mode $m$. 
For Fermionic modes, one has $n_{\max,  m}  =1$, while for Bosonic modes  one has $n_{\max, m}  = \infty$.   To be fully general, we allow  $n_{\max,m}$ to  be any number in  $\N  \cup \{\infty\}$, and possibly even to depend on $m$.

For each mode $m$, we assume that
\begin{enumerate}
\item  the zero-particle subspace $\spc H_{0,m}$ is one-dimensional, meaning that there is a unique vacuum state, hereafter denoted as $|0,m\>$, and 
\item the one-particle subspace  $\spc H_{1,m}$ has dimension $d$,  independently of $m$. 
\end{enumerate}
The second assumption guarantees that we can interpret the one-particle subspace as representing ``a $d$-dimensional quantum system travelling on path $m$.''   

Suppose that the evolution of mode $m$ is described by a quantum channel $\widetilde{\map E}^{(m)}$ that preserves the number of particles. The  Kraus operators of any such channel must have the block-diagonal form  $\widetilde E_{i}^{(m)}   =  \bigoplus_{n=0}^{n_{\max,  m}}   \widetilde E_{i,n}^{(m)}$, where   $(\widetilde E_{i,n}^{(m)})$  are operators operator acting on the $n$-particle subspace, and satisfying the normalization condition 
\begin{align}\label{normj}
\sum_i  \widetilde E_{i,n}^{(m) \, \dag}    \widetilde E_{i,n}^{(m)}   =  I_n^{(m)} \,,
\end{align} where $I_n^{(m)}$ is the identity operator on the $n$-particle subspace of mode $m$ \cite{chiribella2017optimal}.   In particular, since the zero-particle subspace is one-dimensional, the operators $E_{i,0}^{(m)}$ are complex numbers, called the {\em vacuum amplitudes} of channel $\widetilde{\map E}^{(m)}$, and hereafter denoted as $\alpha_i^{(m)}:  = \widetilde E_{i,0}^{(m)}$.   Note that Eq. (\ref{normj}) becomes  the normalization condition 
\begin{align}\label{amplinorm}
\sum_i  |\alpha_i^{(m)}|^2   = 1\,  .
\end{align}    

On the one-particle subspace, the channel $\widetilde{\map E}^{(m)}$ acts as a quantum channel $\map E^{(m)}$ with Kraus operators $E^{(m)}_i  :   =  \widetilde E_{i,1}^{(m)}$.    We call channel  $\widetilde{\map E}^{(m)}$  an {\em extension} of  channel  $\map E^{(m)}$.   

Now, consider the situation of a single particle propagating in a coherent superposition of $N$ paths.   The state space of the single particle is the one-particle subspace of the $N$ modes associated to the $N$ paths.  A generic state in the one-particle subspace is of  the form 
\begin{align}
|\Psi\>     =  \sum_{  m=0}^{N-1}    \,  c_m 
|0,0  \>  \otimes   \cdots \otimes |0,m-1\>  \otimes   |\psi_{m}  \> \otimes  |0,m+1\>  \otimes \cdots \otimes |0, N-1\>\, ,     
\end{align}
that is, it is a linear combination of product states where one mode is in a one-particle state and all the other modes are in the vacuum.    The one-particle subspace can be equivalently represented as a bipartite system, whose subsystems are   an internal degree of freedom of the particle (denoted by $S$),  and the particle's path (denoted by $C$, in analogy to the control system in the quantum SWITCH).  Explicitly,  the one-particle states can be  written as 
\begin{align}
|\Psi\>     =  \sum_{  m=0}^{N-1}    \,  c_m 
    |\psi_{m}  \>_S \otimes |m\>_{C} \, ,     
\end{align}
 where we introduced   the notation  $|\psi_m\>_S  \otimes |m\>_C  :  =  |0,m  \>  \otimes   \cdots \otimes |0,m-1\>  \otimes   |\psi_{m}  \> \otimes  |0,m+1\>  \otimes \cdots \otimes |0, N-1\>$, and associated the orthonormal vectors $(|m\>)_{m=0}^{N-1}$ to  the $N$ possible paths that the particle can traverse.

 Assuming that the $N$ modes evolve independently under the channels $(\widetilde{\map E}^{(m)} )_{m=0}^{N-1}$,  the evolution of the single particle    is simply the restriction of the product channel $\widetilde{\map E}^{(0)} \otimes  \widetilde{\map E}^{(1)}   \otimes \cdots \otimes \widetilde{\map E}^{(N-1)}$ to the one-particle subspace. 
 We denote the one-particle restriction by  $\map R  (\widetilde {\map E}^{(0)},  \widetilde {\map E}^{(1)}, \dots,  \widetilde {\map E}^{(N-1)})$, and its   Kraus operators by  
\begin{align}
\nonumber 
&R_{i_0 i_1 \dots i_{N-1}} \\
\label{superposition0}
&  =    \sum_{m=0}^{N-1}  \,     \alpha^{(0)}_{i_0}    |0,1\>\<0,1|   \otimes  \cdots\otimes   \alpha^{(m-1)}_{i_{m-1}}     \,  |0,m-1\>\<0,m-1|\otimes E^{(m)}_{i_m}  \otimes \alpha^{(m+1)}_{i_{m+1}}  \,  |0,m+1\>\<0,m+1|  \otimes \cdots  \otimes   \alpha^{(N-1)}_{i_{N-1}}  |0,N-1\>\<0,N-1|  \, ,            
\end{align}
or equivalently, 
\begin{align}\label{superposition}
R_{i_0 i_1 \dots i_{N-1}}  =    \sum_{m=0}^{N-1} \,        \alpha^{(0)}_{i_0}   \cdots  \alpha^{(m-1)}_{i_{m-1}}  E^{(m)}_{i_m} \alpha^{(m+1)}_{i_{m+1}}  \cdots    \alpha^{(N-1)}_{i_{N-1}}  \otimes |m\>\<m|_C  \, .          
\end{align}
Note that the one-particle restriction depends only on the  one-particle channels $(\map E^{(m)})_{m=0}^{N-1}$  and on the vacuum amplitudes $(\alpha^{(m)})_{m=0}^{N-1}$.    Hence, we can without loss of generality assume that the maximum number of particles  is $n_{\max, m}  = 1$ for every mode $m$.    This assumption will help simplifying the characterization of the possible extensions of the original channels $(\map E^{(m)})_{m=0}^{N-1}$.

In the following, we make the standard assumption that the path is initialized in a fixed state $\omega$, independent of the state of the internal degree of freedom $S$ \cite{Gisin05, Oi03, aaberg2004subspace, Abbott2018, Giulio18, Hler20}. Then, the communication between the sender and the receiver is described by  the effective channel   $\map R_\omega   ( \widetilde{\map E}^{(0)} ,\dots, \widetilde{\map E}^{(N-1)})$ defined by the relation
\begin{align}
\map R_\omega   ( \widetilde{\map E}^{(0)} ,\dots, \widetilde{\map E}^{(N-1)})  \,   (\rho) :  = \map R ( \widetilde{\map E}^{(0)} ,\dots, \widetilde{\map E}^{(N-1)})  \,   (\rho \otimes \omega) \, . 
\end{align}   
We call the channel  $\map R_\omega   ( \widetilde{\map E}^{(0)} ,\dots, \widetilde{\map E}^{(N-1)})$   a {\em coherent superposition   of the channels $(\map E^{(m)})_{m=0}^{N-1}$}, or simply, the {\em superposition channel}.  

\section{Proof of Theorem \ref{theo:path}}\label{app:path}

Here we show that, if all the channels $(\map E^{(m)})_{m=0}^{N-1}$  are noisy, then it is impossible to find  extensions $(\widetilde{\map E}^{(m)})_{m=0}^{N-1}$ and a state $\omega$ such that the superposition channel  $\map R_\omega   ( \widetilde{\map E}^{(0)} ,\dots, \widetilde{\map E}^{(N-1)})$ is correctable.  

First, note that we can assume without loss of generality that the state $\omega$ is pure. Indeed, the   superposition channel $\map R_\omega  ( \widetilde{\map E}^{(0)} ,\dots, \widetilde{\map E}^{(N-1)})$  depends linearly on the state $\omega$, and one has $\map R_{p \, \omega+ (1-p)\,  \omega' } ( \widetilde{\map E}^{(0)} ,\dots, \widetilde{\map E}^{(N-1)})   =  p\,  \map R_\omega  ( \widetilde{\map E}^{(0)} ,\dots, \widetilde{\map E}^{(N-1)})  +  (1-p) \,  \map R_{\omega'}  ( \widetilde{\map E}^{(0)} ,\dots, \widetilde{\map E}^{(N-1)})$.  Since the convex combination of two channels is correctable if and only if each channel is correctable, if the superposition channel is correctable for a mixture $p\, \omega + (1-p) \, \omega'$,  then it must be correctable for each of the two states $\omega$ and $\omega'$.  Hence, we can without loss of generality assume that the state $\omega$ is pure, namely  $\omega=  |\phi\>\<\phi|$ for some unit vector 
\begin{align}\label{state}
|\phi\>  =  \sum_{m=0}^{N-1}  \,  c_m  \,  |m\>  \, .
\end{align}
With this choice,  the superposition channel  $\map R_\omega   ( \widetilde{\map E}^{(0)} ,\dots, \widetilde{\map E}^{(N-1)})$ has   Kraus operators 
\begin{align}
\nonumber R_{\omega,i_0 i_1 \dots i_{N-1}} & :=      R_{i_0 i_1 \dots i_{N-1}}     \,   (I_S  \otimes  |\phi\>_C)   \\
 & =  \sum_{m=0}^{N-1} \,    c_j\,     \alpha^{(0)}_{i_0}   \cdots  \alpha^{(m-1)}_{i_{m-1}}  E^{(m)}_{i_m} \alpha^{(m+1)}_{i_{m+1}}  \cdots    \alpha^{(N-1)}_{i_{N-1}}  \otimes |m\>_C  \, , 
\end{align}
where the notation  $A \otimes |\phi\>_C$  denotes the linear  operator from $\spc H_S$ to $\spc H_{S} \otimes \spc H_C$ defined by the relation  $\left(A  \otimes  |\phi\>_C\right)  \, |\psi\>_S : =  (A|\psi\> )_S\otimes |\phi\>_C  \,, \forall |\psi\>  \in  \spc H_S$.

Second, note that we can assume without loss of generality that the unit vector  $|\phi\>$ satisfies the condition $c_m\not  = 0$ for every path $m$.  Indeed, if any of the paths has zero amplitude, we can simply remove it from the list of allowed paths, and focus our attention on the remaining paths.

Third, note that, for every given $m$,  the extensions of a given channel $\map E^{(m)}$  form a convex set: if $\widetilde{\map E}^{(m)}$ and $\widetilde{\map E}^{(m)\prime}$ are two extensions of $\map E^{(m)}$, then also the channel $   p \widetilde{\map E}^{(m)} +  (1-p) \,  \widetilde{\map E}^{(m)   \prime}$ is an extension of $\map E^{(m)}$, for every probability $p\in  [0,1]$.  
 Without loss of generality, the  extension $\widetilde{\map E}^{(m)}$ can be taken to be an extreme point of the convex set.  Indeed, the   superposition channel $\map R_\omega   ( \widetilde{\map E}^{(0)} ,\dots, \widetilde{\map E}^{(N-1)})$  depends linearly on the extensions $(\widetilde{\map E}^{(m)})_{m=0}^{N-1}$, and convex combinations of the form   $p \widetilde{\map E}^{(m)} +  (1-p) \,  \widetilde{\map E}^{(m)   \prime}$ result into convex combinations of the form $   p  \map R_\omega   ( \widetilde{\map E}^{(0)} , \dots, \widetilde{\map E}^{(m)},  \dots,\widetilde{\map E}^{(N-1)})  +  (1-p) \,  \map R_\omega  ( \widetilde{\map E}^{(0)} ,\dots,\widetilde{\map E}^{(m)\prime}, \dots ,\widetilde{\map E}^{(N-1)})$.  Since the convex combination of two channels is correctable only if each channel is correctable (Lemma \ref{lem:pure}), this argument allows us to restrict our attention to the case where each channel $\widetilde {\map E}^{(m)}$ is an extreme point.

An important property of the extreme points of the set of extensions of a given channel was established  in \cite{Giulio18}  (Appendix D, Proposition 5): for an extreme extension $\widetilde{\map E}^{(m)}$ with maximum number of particles $n_{\max, m}  = 1$, every Kraus representation consisting of linearly independent operators $(\widetilde E^{(m)}_i )_{i=1}^r$ must satisfy the condition  
\begin{align}\label{nonzerokraus}
E^{(m)}_i \not  =  0  \qquad \forall   i\in  \{1,\dots,  r\} \, .
\end{align} 
   Using this fact, we will now show that the superposition channel  $\map R_\omega   ( \widetilde{\map E}^{(0)} ,\dots, \widetilde{\map E}^{(N-1)})$ is not correctable for any choice of extreme extensions, and therefore for every choice of extensions.

The superposition  channel $\map R_\omega   ( \widetilde{\map E}^{(0)} ,\dots, \widetilde{\map E}^{(N-1)})$   has a $d$-dimensional input and a $(Nd)$-dimensional output.  In order to be correctable on the whole  input space, $\map R_\omega   ( \widetilde{\map E}^{(0)} ,\dots, \widetilde{\map E}^{(N-1)})$  can have at most $N$ linearly independent Kraus operators (see e.g. \cite{Giulio11}). We now show that, if all the channels $(\map E^{(m)})_{m=0}^{N-1}$ are noisy, then  a superposition of $N$ independent noisy channels must have at least $N+1$ linearly independent Kraus operators, and therefore cannot be correctable.

For every given $m$, the normalization of the amplitudes  (\ref{amplinorm}) guarantees that there exists one value $p_m$ such that $\alpha^{(m)}_{p_m}  \not  = 0$.   Moreover, since the extension $\widetilde {\map E}^{(m)}$ is extreme, also the Kraus operator $E^{(m)}_{p_m}$ is non-zero, thanks to Eq. (\ref{nonzerokraus}).  

Since each channel  $\map E^{(m)}$   is noisy, there exists at least one value $q_m$ such that the   Kraus operators $E^{(m)}_{p_m}$ and $E^{(m)}_{q_m}$ are linearly independent.  
Hence, the superposition channel $\map R_\omega   ( \widetilde{\map E}^{(1)} ,\dots, \widetilde{\map E}^{(N-1)})$ has at least   $N+1$ Kraus operators 
\begin{align}
\nonumber   R_{\omega,0}  &:   = R_{\omega,q_0 p_1  \cdots  p_{N-1}}      \\
\nonumber   R_{\omega,1}   &: = R_{\omega, p_0 q_1 p_2 \cdots  p_{N-1}  }  \\
\nonumber &~~\vdots \\
\nonumber  R_{\omega, N-1}   &:  =  R_{\omega, p_0    \cdots   p_{N-2} q_{N-1} }   \\ 
   R_{\omega, N}    &:  =  R_{\omega,p_0 \cdots    p_{N-1}}  \,.    
\end{align}
These operators are linearly independent: if $\{\lambda_k\}_{k=0}^N$ are coefficients such that $\sum_k  \,\lambda_k \, R_{\omega,k}  =  0$, then we can multiply by $(I_S\otimes \< m|_C )$  on the left, obtaining the relation 
\begin{align}
\lambda_m  \,  c_m\,  \alpha^{(0)}_{p_0}  \cdots   \alpha^{(m-1)}_{p_m}        \alpha^{(m+1)}_{p_m} \cdots    \alpha^{(N-1)}_{p_{N-1}}   \,   E^{(m)}_{q_m}  +  \widetilde \lambda_m  \,  E^{(m)}_{p_m}  = 0 \, ,
\end{align}
where $\widetilde \lambda_m$ is a suitable constant.   Since the operators $E^{(m)}_{  q_m}$ and $E^{(m)}_{p_m}$ are linearly independent, we must have   $\lambda_m  \,   c_m \,  \alpha^{(0)}_{p_0}  \cdots   \alpha^{(m-1)}_{p_m}        \alpha^{(m+1)}_{p_m} \cdots    \alpha^{(N-1)}_{p_{N-1}}   =  0$, which in turn implies $\lambda_m=  0$ because  each $c_m$ is non-zero,  and all   the coefficients $\alpha^{(m)}_{p_m}$ are  non-zero by construction. 
In this way, we obtain $\lambda_m= 0$ for every $m \in  \{ 0,\dots,  N-1\}$.  Hence,  the condition $\sum_k \lambda_k\,  R_{\omega, k} =0$ reduces to $\lambda_N  \, R_{\omega,N}  =  0$. Since   the operator $R_{\omega, N}   =  \sum_{m=1}^{N-1} \,    c_m\,     \alpha^{(0)}_{p_1}   \cdots  \alpha^{(m-1)}_{p_{m-1}}  E^{(m)}_{p_m} \alpha^{(m+1)}_{p_{m+1}}  \cdots    \alpha^{(N-1)}_{p_{N-1}}  \otimes |m\>_C  $ is non-zero by construction, we conclude that  $\lambda_N  =  0$.   In conclusion, all the coefficients $(\lambda_k)_{k=0}^N$ must be zero.    

In summary, the superposition channel  $\map R_\omega   ( \widetilde{\map E}^{(0)} ,\dots, \widetilde{\map E}^{(N-1)})$   has at least $N+1$ linearly independent operators, and therefore it cannot be corrected.   \qed

\section{Proof of Corollary \ref{cor:finitedistance}}\label{app:finitedistance}

Let ${\sf Chan}  (S, SC)$ be the set of all quantum channels from $S$ to $SC$.   By assumption, both the dimension of $S$ (equal to $d$) and the dimension of $C$ (equal to the number of paths $N$) are finite.  Hence, the  set of channels ${\sf Chan}  (S,SC)$  is a finite-dimensional compact set. 

Recall that  all norms on finite dimensions are equivalent.  In the following we will denote by $\|  \cdot \|$ a generic norm on the set of quantum channels   from $S$ to $SC$.  

  Let ${\sf Correctable} (S, SC)$ be the set of correctable channels from $S$ to $SC$, that is, the set of channels $\map C$ for which there exists another channel $\map C'$ such that $\map C'\circ \map C  = \map I$.  Note that the set  ${\sf Correctable} (S, SC)$ is compact, due to the compactness of ${\sf Chan}  (S, SC)$.  
  
     For a given channel $\map R \in  {\sf Chan}  (S, SC) $ define the distance between $\map R$ and the set of correctable channels: 
  \begin{align}\label{delta}
  \delta(\map R) :  =   \inf_{\map C  \in  {\sf Correctable} (S, SC)}    \left \|   \map R - \map  C \right\|  \, .    
  \end{align}

Now, let ${\sf Chan}  (S)$ be the set of channels from system $S$ to itself.  For a given channel $\map E  \in  {\sf Chan}  (S)$, let $\widetilde {\map E}$ be an extension of channel $\map E$, acting on the original system $S$ and on the vacuum.  Define ${\sf Ext}  (\map E)$ be the set of all such  extensions.   Again, since the dimension of the input system is finite, the set ${\sf Ext}  (\map E)$ is a finite-dimensional compact set.  

For a list of channels $(\map E^{(0)}, \dots,  \map E^{(N-1)})$, let us define the distance between a generic superposition of the channels $(\map E^{(0)}, \dots,  \map E^{(N-1)})$ and the set of correctable channels:
\begin{align}\label{inf}
\delta_{*}  (\map E^{(0)}, \dots,  \map E^{(N-1)})  :  =  \inf_{\omega  \in  D  (\spc H_C)}   \inf_{  \widetilde {\map E}^{(0)}  \in  {\sf Ext}  (\map E^{(0)})  }  \,   \cdots  \inf_{  \widetilde {\map E}^{(N-1)}  \in  {\sf Ext}  (\map E^{(N-1)})  }\,  
\delta  (\map S_\omega (\widetilde{\map E}^{(0)}, \dots,  \widetilde{\map E}^{(N-1)})) \, .
\end{align}

The claim of Corollary \ref{cor:finitedistance} is that $\delta_*  (\map E^{(0)}, \dots,  \map E^{(N-1)})$ is a strictly positive number.  This claim follows from the fact that the function $\delta$ in Eq. (\ref{delta}) is continuous, and  all the sets in the right-hand side of Eq. (\ref{inf}) are compact sets, which implies that  all the infima are actually minima. In other words, there exists a state $\omega$, and extensions  $(  \widetilde  {\map E}^{(0)},  \dots,  \widetilde {\map E}^{(N-1)})$ such that  $\delta_*  (\map E^{(0)}, \dots,  \map E^{(N-1)})   =  \delta  (\map S_\omega (\widetilde{\map E}^{(0)}, \dots,  \widetilde{\map E}^{(N-1)}))$.  By  Theorem \ref{theo:path}, one then has $\delta  (\map S_\omega (\widetilde{\map E}^{(0)}, \dots,  \widetilde{\map E}^{(N-1)}))  >  0$. \qed

\bibliography{PerfectActivation_arxiv}

\begin{thebibliography}{59}%
\makeatletter
\providecommand \@ifxundefined [1]{%
 \@ifx{#1\undefined}
}%
\providecommand \@ifnum [1]{%
 \ifnum #1\expandafter \@firstoftwo
 \else \expandafter \@secondoftwo
 \fi
}%
\providecommand \@ifx [1]{%
 \ifx #1\expandafter \@firstoftwo
 \else \expandafter \@secondoftwo
 \fi
}%
\providecommand \natexlab [1]{#1}%
\providecommand \enquote  [1]{``#1''}%
\providecommand \bibnamefont  [1]{#1}%
\providecommand \bibfnamefont [1]{#1}%
\providecommand \citenamefont [1]{#1}%
\providecommand \href@noop [0]{\@secondoftwo}%
\providecommand \href [0]{\begingroup \@sanitize@url \@href}%
\providecommand \@href[1]{\@@startlink{#1}\@@href}%
\providecommand \@@href[1]{\endgroup#1\@@endlink}%
\providecommand \@sanitize@url [0]{\catcode `\\12\catcode `\$12\catcode
  `\&12\catcode `\#12\catcode `\^12\catcode `\_12\catcode `\%12\relax}%
\providecommand \@@startlink[1]{}%
\providecommand \@@endlink[0]{}%
\providecommand \url  [0]{\begingroup\@sanitize@url \@url }%
\providecommand \@url [1]{\endgroup\@href {#1}{\urlprefix }}%
\providecommand \urlprefix  [0]{URL }%
\providecommand \Eprint [0]{\href }%
\providecommand \doibase [0]{http://dx.doi.org/}%
\providecommand \selectlanguage [0]{\@gobble}%
\providecommand \bibinfo  [0]{\@secondoftwo}%
\providecommand \bibfield  [0]{\@secondoftwo}%
\providecommand \translation [1]{[#1]}%
\providecommand \BibitemOpen [0]{}%
\providecommand \bibitemStop [0]{}%
\providecommand \bibitemNoStop [0]{.\EOS\space}%
\providecommand \EOS [0]{\spacefactor3000\relax}%
\providecommand \BibitemShut  [1]{\csname bibitem#1\endcsname}%
\let\auto@bib@innerbib\@empty
\bibitem [{\citenamefont {Shannon}(1948)}]{Shannon48}%
  \BibitemOpen
  \bibfield  {author} {\bibinfo {author} {\bibfnamefont {Claude~E}\
  \bibnamefont {Shannon}},\ }\bibfield  {title} {\enquote {\bibinfo {title} {A
  mathematical theory of communication},}\ }\href@noop {} {\bibfield  {journal}
  {\bibinfo  {journal} {The Bell System Technical Journal}\ }\textbf {\bibinfo
  {volume} {27}},\ \bibinfo {pages} {379--423} (\bibinfo {year}
  {1948})}\BibitemShut {NoStop}%
\bibitem [{\citenamefont {Bennett}\ and\ \citenamefont
  {Brassard}(1984)}]{Bennett85}%
  \BibitemOpen
  \bibfield  {author} {\bibinfo {author} {\bibfnamefont {CH}~\bibnamefont
  {Bennett}}\ and\ \bibinfo {author} {\bibfnamefont {G}~\bibnamefont
  {Brassard}},\ }\bibfield  {title} {\enquote {\bibinfo {title} {Quantum
  cryptography: Public key distribution and coin tossing},}\ }\href@noop {}
  {\bibfield  {journal} {\bibinfo  {journal} {Proceedings of the International
  Conference on Computers, Systems and Signal Processing, Bangalore, India}\ ,\
  \bibinfo {pages} {175--179}} (\bibinfo {year} {1984})}\BibitemShut {NoStop}%
\bibitem [{\citenamefont {Ekert}(1991)}]{Ekert91}%
  \BibitemOpen
  \bibfield  {author} {\bibinfo {author} {\bibfnamefont {Artur~K}\ \bibnamefont
  {Ekert}},\ }\bibfield  {title} {\enquote {\bibinfo {title} {Quantum
  cryptography based on bell’s theorem},}\ }\href@noop {} {\bibfield
  {journal} {\bibinfo  {journal} {Physical Review Letters}\ }\textbf {\bibinfo
  {volume} {67}},\ \bibinfo {pages} {661} (\bibinfo {year} {1991})}\BibitemShut
  {NoStop}%
\bibitem [{\citenamefont {Bennett}\ and\ \citenamefont
  {Wiesner}(1992)}]{Bennett92}%
  \BibitemOpen
  \bibfield  {author} {\bibinfo {author} {\bibfnamefont {Charles~H}\
  \bibnamefont {Bennett}}\ and\ \bibinfo {author} {\bibfnamefont {Stephen~J}\
  \bibnamefont {Wiesner}},\ }\bibfield  {title} {\enquote {\bibinfo {title}
  {Communication via one-and two-particle operators on einstein-podolsky-rosen
  states},}\ }\href@noop {} {\bibfield  {journal} {\bibinfo  {journal}
  {Physical Review Letters}\ }\textbf {\bibinfo {volume} {69}},\ \bibinfo
  {pages} {2881} (\bibinfo {year} {1992})}\BibitemShut {NoStop}%
\bibitem [{\citenamefont {Holevo}(1998)}]{Holevo98}%
  \BibitemOpen
  \bibfield  {author} {\bibinfo {author} {\bibfnamefont {Alexander~S}\
  \bibnamefont {Holevo}},\ }\bibfield  {title} {\enquote {\bibinfo {title} {The
  capacity of the quantum channel with general signal states},}\ }\href@noop {}
  {\bibfield  {journal} {\bibinfo  {journal} {IEEE Transactions on Information
  Theory}\ }\textbf {\bibinfo {volume} {44}},\ \bibinfo {pages} {269--273}
  (\bibinfo {year} {1998})}\BibitemShut {NoStop}%
\bibitem [{\citenamefont {Schumacher}\ and\ \citenamefont
  {Westmoreland}(1997)}]{Schumacher97}%
  \BibitemOpen
  \bibfield  {author} {\bibinfo {author} {\bibfnamefont {Benjamin}\
  \bibnamefont {Schumacher}}\ and\ \bibinfo {author} {\bibfnamefont
  {Michael~D}\ \bibnamefont {Westmoreland}},\ }\bibfield  {title} {\enquote
  {\bibinfo {title} {Sending classical information via noisy quantum
  channels},}\ }\href@noop {} {\bibfield  {journal} {\bibinfo  {journal}
  {Physical Review A}\ }\textbf {\bibinfo {volume} {56}},\ \bibinfo {pages}
  {131} (\bibinfo {year} {1997})}\BibitemShut {NoStop}%
\bibitem [{\citenamefont {Lloyd}(1997)}]{Lloyd97}%
  \BibitemOpen
  \bibfield  {author} {\bibinfo {author} {\bibfnamefont {Seth}\ \bibnamefont
  {Lloyd}},\ }\bibfield  {title} {\enquote {\bibinfo {title} {Capacity of the
  noisy quantum channel},}\ }\href@noop {} {\bibfield  {journal} {\bibinfo
  {journal} {Physical Review A}\ }\textbf {\bibinfo {volume} {55}},\ \bibinfo
  {pages} {1613} (\bibinfo {year} {1997})}\BibitemShut {NoStop}%
\bibitem [{\citenamefont {Shor}(2002)}]{Shor02}%
  \BibitemOpen
  \bibfield  {author} {\bibinfo {author} {\bibfnamefont {PW}~\bibnamefont
  {Shor}},\ }\bibfield  {title} {\enquote {\bibinfo {title} {talk at msri
  workshop on quantum computation},}\ }\href@noop {} {\bibfield  {journal}
  {\bibinfo  {journal} {available online under http://www. msri.
  org/publications/ln/msri/2002/quantumcrypto/shor/1}\ } (\bibinfo {year}
  {2002})}\BibitemShut {NoStop}%
\bibitem [{\citenamefont {Devetak}(2005)}]{Devetak05}%
  \BibitemOpen
  \bibfield  {author} {\bibinfo {author} {\bibfnamefont {Igor}\ \bibnamefont
  {Devetak}},\ }\bibfield  {title} {\enquote {\bibinfo {title} {The private
  classical capacity and quantum capacity of a quantum channel},}\ }\href@noop
  {} {\bibfield  {journal} {\bibinfo  {journal} {IEEE Transactions on
  Information Theory}\ }\textbf {\bibinfo {volume} {51}},\ \bibinfo {pages}
  {44--55} (\bibinfo {year} {2005})}\BibitemShut {NoStop}%
\bibitem [{\citenamefont {Bennett}\ \emph {et~al.}(1996)\citenamefont
  {Bennett}, \citenamefont {DiVincenzo}, \citenamefont {Smolin},\ and\
  \citenamefont {Wootters}}]{Bennett96}%
  \BibitemOpen
  \bibfield  {author} {\bibinfo {author} {\bibfnamefont {Charles~H}\
  \bibnamefont {Bennett}}, \bibinfo {author} {\bibfnamefont {David~P}\
  \bibnamefont {DiVincenzo}}, \bibinfo {author} {\bibfnamefont {John~A}\
  \bibnamefont {Smolin}}, \ and\ \bibinfo {author} {\bibfnamefont {William~K}\
  \bibnamefont {Wootters}},\ }\bibfield  {title} {\enquote {\bibinfo {title}
  {Mixed-state entanglement and quantum error correction},}\ }\href@noop {}
  {\bibfield  {journal} {\bibinfo  {journal} {Physical Review A}\ }\textbf
  {\bibinfo {volume} {54}},\ \bibinfo {pages} {3824} (\bibinfo {year}
  {1996})}\BibitemShut {NoStop}%
\bibitem [{\citenamefont {Bennett}\ \emph {et~al.}(1999)\citenamefont
  {Bennett}, \citenamefont {Shor}, \citenamefont {Smolin},\ and\ \citenamefont
  {Thapliyal}}]{Bennett99}%
  \BibitemOpen
  \bibfield  {author} {\bibinfo {author} {\bibfnamefont {Charles~H}\
  \bibnamefont {Bennett}}, \bibinfo {author} {\bibfnamefont {Peter~W}\
  \bibnamefont {Shor}}, \bibinfo {author} {\bibfnamefont {John~A}\ \bibnamefont
  {Smolin}}, \ and\ \bibinfo {author} {\bibfnamefont {Ashish~V}\ \bibnamefont
  {Thapliyal}},\ }\bibfield  {title} {\enquote {\bibinfo {title}
  {Entanglement-assisted classical capacity of noisy quantum channels},}\
  }\href@noop {} {\bibfield  {journal} {\bibinfo  {journal} {Physical Review
  Letters}\ }\textbf {\bibinfo {volume} {83}},\ \bibinfo {pages} {3081}
  (\bibinfo {year} {1999})}\BibitemShut {NoStop}%
\bibitem [{\citenamefont {Nielsen}\ and\ \citenamefont
  {Chuang}(2000)}]{Chuang00}%
  \BibitemOpen
  \bibfield  {author} {\bibinfo {author} {\bibfnamefont {Michael~A}\
  \bibnamefont {Nielsen}}\ and\ \bibinfo {author} {\bibfnamefont {Isaac}\
  \bibnamefont {Chuang}},\ }\href@noop {} {\emph {\bibinfo {title} {Quantum
  computation and quantum information}}}\ (\bibinfo  {publisher} {Cambridge
  University Press},\ \bibinfo {year} {2000})\BibitemShut {NoStop}%
\bibitem [{\citenamefont {Wilde}(2013)}]{Wilde13}%
  \BibitemOpen
  \bibfield  {author} {\bibinfo {author} {\bibfnamefont {Mark~M}\ \bibnamefont
  {Wilde}},\ }\href@noop {} {\emph {\bibinfo {title} {Quantum information
  theory}}}\ (\bibinfo  {publisher} {Cambridge University Press},\ \bibinfo
  {year} {2013})\BibitemShut {NoStop}%
\bibitem [{\citenamefont {Aharonov}\ \emph {et~al.}(1990)\citenamefont
  {Aharonov}, \citenamefont {Anandan}, \citenamefont {Popescu},\ and\
  \citenamefont {Vaidman}}]{Aharonov90}%
  \BibitemOpen
  \bibfield  {author} {\bibinfo {author} {\bibfnamefont {Yakir}\ \bibnamefont
  {Aharonov}}, \bibinfo {author} {\bibfnamefont {Jeeva}\ \bibnamefont
  {Anandan}}, \bibinfo {author} {\bibfnamefont {Sandu}\ \bibnamefont
  {Popescu}}, \ and\ \bibinfo {author} {\bibfnamefont {Lev}\ \bibnamefont
  {Vaidman}},\ }\bibfield  {title} {\enquote {\bibinfo {title} {Superpositions
  of time evolutions of a quantum system and a quantum time-translation
  machine},}\ }\href@noop {} {\bibfield  {journal} {\bibinfo  {journal}
  {Physical Review Letters}\ }\textbf {\bibinfo {volume} {64}},\ \bibinfo
  {pages} {2965} (\bibinfo {year} {1990})}\BibitemShut {NoStop}%
\bibitem [{\citenamefont {Oi}(2003)}]{Oi03}%
  \BibitemOpen
  \bibfield  {author} {\bibinfo {author} {\bibfnamefont {Daniel~KL}\
  \bibnamefont {Oi}},\ }\bibfield  {title} {\enquote {\bibinfo {title}
  {Interference of quantum channels},}\ }\href@noop {} {\bibfield  {journal}
  {\bibinfo  {journal} {Physical Review Letters}\ }\textbf {\bibinfo {volume}
  {91}},\ \bibinfo {pages} {067902} (\bibinfo {year} {2003})}\BibitemShut
  {NoStop}%
\bibitem [{\citenamefont {{\AA}berg}(2004)}]{aaberg2004subspace}%
  \BibitemOpen
  \bibfield  {author} {\bibinfo {author} {\bibfnamefont {Johan}\ \bibnamefont
  {{\AA}berg}},\ }\bibfield  {title} {\enquote {\bibinfo {title} {Subspace
  preservation, subspace locality, and gluing of completely positive maps},}\
  }\href@noop {} {\bibfield  {journal} {\bibinfo  {journal} {Annals of
  Physics}\ }\textbf {\bibinfo {volume} {313}},\ \bibinfo {pages} {326--367}
  (\bibinfo {year} {2004})}\BibitemShut {NoStop}%
\bibitem [{\citenamefont {Gisin}\ \emph {et~al.}(2005)\citenamefont {Gisin},
  \citenamefont {Linden}, \citenamefont {Massar},\ and\ \citenamefont
  {Popescu}}]{Gisin05}%
  \BibitemOpen
  \bibfield  {author} {\bibinfo {author} {\bibfnamefont {N.}~\bibnamefont
  {Gisin}}, \bibinfo {author} {\bibfnamefont {N.}~\bibnamefont {Linden}},
  \bibinfo {author} {\bibfnamefont {S.}~\bibnamefont {Massar}}, \ and\ \bibinfo
  {author} {\bibfnamefont {S.}~\bibnamefont {Popescu}},\ }\bibfield  {title}
  {\enquote {\bibinfo {title} {Error filtration and entanglement purification
  for quantum communication},}\ }\href@noop {} {\bibfield  {journal} {\bibinfo
  {journal} {Phys. Rev. A}\ }\textbf {\bibinfo {volume} {72}},\ \bibinfo
  {pages} {012338} (\bibinfo {year} {2005})}\BibitemShut {NoStop}%
\bibitem [{\citenamefont {Chiribella}\ \emph
  {et~al.}(2009{\natexlab{a}})\citenamefont {Chiribella}, \citenamefont
  {D’Ariano}, \citenamefont {Perinotti},\ and\ \citenamefont
  {Valiron}}]{chiribella2009beyond}%
  \BibitemOpen
  \bibfield  {author} {\bibinfo {author} {\bibfnamefont {G}~\bibnamefont
  {Chiribella}}, \bibinfo {author} {\bibfnamefont {GM}~\bibnamefont
  {D’Ariano}}, \bibinfo {author} {\bibfnamefont {P}~\bibnamefont
  {Perinotti}}, \ and\ \bibinfo {author} {\bibfnamefont {B}~\bibnamefont
  {Valiron}},\ }\bibfield  {title} {\enquote {\bibinfo {title} {Beyond quantum
  computers},}\ }\href@noop {} {\bibfield  {journal} {\bibinfo  {journal}
  {arXiv preprint arXiv:0912.0195}\ } (\bibinfo {year}
  {2009}{\natexlab{a}})}\BibitemShut {NoStop}%
\bibitem [{\citenamefont {Chiribella}\ \emph {et~al.}(2013)\citenamefont
  {Chiribella}, \citenamefont {D'Ariano}, \citenamefont {Perinotti},\ and\
  \citenamefont {Valiron}}]{Chiribella13}%
  \BibitemOpen
  \bibfield  {author} {\bibinfo {author} {\bibfnamefont {Giulio}\ \bibnamefont
  {Chiribella}}, \bibinfo {author} {\bibfnamefont {Giacomo~Mauro}\ \bibnamefont
  {D'Ariano}}, \bibinfo {author} {\bibfnamefont {Paolo}\ \bibnamefont
  {Perinotti}}, \ and\ \bibinfo {author} {\bibfnamefont {Benoit}\ \bibnamefont
  {Valiron}},\ }\bibfield  {title} {\enquote {\bibinfo {title} {Quantum
  computations without definite causal structure},}\ }\href@noop {} {\bibfield
  {journal} {\bibinfo  {journal} {Phys. Rev. A}\ }\textbf {\bibinfo {volume}
  {88}},\ \bibinfo {pages} {022318} (\bibinfo {year} {2013})}\BibitemShut
  {NoStop}%
\bibitem [{\citenamefont {Oreshkov}\ \emph {et~al.}(2012)\citenamefont
  {Oreshkov}, \citenamefont {Costa},\ and\ \citenamefont
  {Brukner}}]{Oreshkov12}%
  \BibitemOpen
  \bibfield  {author} {\bibinfo {author} {\bibfnamefont {Ognyan}\ \bibnamefont
  {Oreshkov}}, \bibinfo {author} {\bibfnamefont {Fabio}\ \bibnamefont {Costa}},
  \ and\ \bibinfo {author} {\bibfnamefont {{\v{C}}aslav}\ \bibnamefont
  {Brukner}},\ }\bibfield  {title} {\enquote {\bibinfo {title} {Quantum
  correlations with no causal order},}\ }\href@noop {} {\bibfield  {journal}
  {\bibinfo  {journal} {Nature Communications}\ }\textbf {\bibinfo {volume}
  {3}},\ \bibinfo {pages} {1--8} (\bibinfo {year} {2012})}\BibitemShut
  {NoStop}%
\bibitem [{\citenamefont {Ara{\'u}jo}\ \emph {et~al.}(2015)\citenamefont
  {Ara{\'u}jo}, \citenamefont {Branciard}, \citenamefont {Costa}, \citenamefont
  {Feix}, \citenamefont {Giarmatzi},\ and\ \citenamefont {Brukner}}]{Araujo15}%
  \BibitemOpen
  \bibfield  {author} {\bibinfo {author} {\bibfnamefont {Mateus}\ \bibnamefont
  {Ara{\'u}jo}}, \bibinfo {author} {\bibfnamefont {Cyril}\ \bibnamefont
  {Branciard}}, \bibinfo {author} {\bibfnamefont {Fabio}\ \bibnamefont
  {Costa}}, \bibinfo {author} {\bibfnamefont {Adrien}\ \bibnamefont {Feix}},
  \bibinfo {author} {\bibfnamefont {Christina}\ \bibnamefont {Giarmatzi}}, \
  and\ \bibinfo {author} {\bibfnamefont {{\v{C}}aslav}\ \bibnamefont
  {Brukner}},\ }\bibfield  {title} {\enquote {\bibinfo {title} {Witnessing
  causal nonseparability},}\ }\href@noop {} {\bibfield  {journal} {\bibinfo
  {journal} {New Journal of Physics}\ }\textbf {\bibinfo {volume} {17}},\
  \bibinfo {pages} {102001} (\bibinfo {year} {2015})}\BibitemShut {NoStop}%
\bibitem [{\citenamefont {Oreshkov}\ and\ \citenamefont
  {Giarmatzi}(2016)}]{Oreshkov16}%
  \BibitemOpen
  \bibfield  {author} {\bibinfo {author} {\bibfnamefont {Ognyan}\ \bibnamefont
  {Oreshkov}}\ and\ \bibinfo {author} {\bibfnamefont {Christina}\ \bibnamefont
  {Giarmatzi}},\ }\bibfield  {title} {\enquote {\bibinfo {title} {Causal and
  causally separable processes},}\ }\href@noop {} {\bibfield  {journal}
  {\bibinfo  {journal} {New Journal of Physics}\ }\textbf {\bibinfo {volume}
  {18}},\ \bibinfo {pages} {093020} (\bibinfo {year} {2016})}\BibitemShut
  {NoStop}%
\bibitem [{\citenamefont {Chiribella}(2012)}]{Chiribella12}%
  \BibitemOpen
  \bibfield  {author} {\bibinfo {author} {\bibfnamefont {Giulio}\ \bibnamefont
  {Chiribella}},\ }\bibfield  {title} {\enquote {\bibinfo {title} {Perfect
  discrimination of no-signalling channels via quantum superposition of causal
  structures},}\ }\href@noop {} {\bibfield  {journal} {\bibinfo  {journal}
  {Physical Review A}\ }\textbf {\bibinfo {volume} {86}},\ \bibinfo {pages}
  {040301} (\bibinfo {year} {2012})}\BibitemShut {NoStop}%
\bibitem [{\citenamefont {Ara{\'u}jo}\ \emph {et~al.}(2014)\citenamefont
  {Ara{\'u}jo}, \citenamefont {Costa},\ and\ \citenamefont
  {Brukner}}]{Araujo14}%
  \BibitemOpen
  \bibfield  {author} {\bibinfo {author} {\bibfnamefont {Mateus}\ \bibnamefont
  {Ara{\'u}jo}}, \bibinfo {author} {\bibfnamefont {Fabio}\ \bibnamefont
  {Costa}}, \ and\ \bibinfo {author} {\bibfnamefont {{\v{C}}aslav}\
  \bibnamefont {Brukner}},\ }\bibfield  {title} {\enquote {\bibinfo {title}
  {Computational advantage from quantum-controlled ordering of gates},}\
  }\href@noop {} {\bibfield  {journal} {\bibinfo  {journal} {Physical Review
  Letters}\ }\textbf {\bibinfo {volume} {113}},\ \bibinfo {pages} {250402}
  (\bibinfo {year} {2014})}\BibitemShut {NoStop}%
\bibitem [{\citenamefont {Gu{\'e}rin}\ \emph {et~al.}(2016)\citenamefont
  {Gu{\'e}rin}, \citenamefont {Feix}, \citenamefont {Ara{\'u}jo},\ and\
  \citenamefont {Brukner}}]{Guerin16}%
  \BibitemOpen
  \bibfield  {author} {\bibinfo {author} {\bibfnamefont {Philippe~Allard}\
  \bibnamefont {Gu{\'e}rin}}, \bibinfo {author} {\bibfnamefont {Adrien}\
  \bibnamefont {Feix}}, \bibinfo {author} {\bibfnamefont {Mateus}\ \bibnamefont
  {Ara{\'u}jo}}, \ and\ \bibinfo {author} {\bibfnamefont {{\v{C}}aslav}\
  \bibnamefont {Brukner}},\ }\bibfield  {title} {\enquote {\bibinfo {title}
  {Exponential communication complexity advantage from quantum superposition of
  the direction of communication},}\ }\href@noop {} {\bibfield  {journal}
  {\bibinfo  {journal} {Physical Review Letters}\ }\textbf {\bibinfo {volume}
  {117}},\ \bibinfo {pages} {100502} (\bibinfo {year} {2016})}\BibitemShut
  {NoStop}%
\bibitem [{\citenamefont {Zhao}\ \emph {et~al.}(2020)\citenamefont {Zhao},
  \citenamefont {Yang},\ and\ \citenamefont {Chiribella}}]{zhao2020quantum}%
  \BibitemOpen
  \bibfield  {author} {\bibinfo {author} {\bibfnamefont {Xiaobin}\ \bibnamefont
  {Zhao}}, \bibinfo {author} {\bibfnamefont {Yuxiang}\ \bibnamefont {Yang}}, \
  and\ \bibinfo {author} {\bibfnamefont {Giulio}\ \bibnamefont {Chiribella}},\
  }\bibfield  {title} {\enquote {\bibinfo {title} {Quantum metrology with
  indefinite causal order},}\ }\href@noop {} {\bibfield  {journal} {\bibinfo
  {journal} {Physical Review Letters}\ }\textbf {\bibinfo {volume} {124}},\
  \bibinfo {pages} {190503} (\bibinfo {year} {2020})}\BibitemShut {NoStop}%
\bibitem [{\citenamefont {Guha}\ \emph {et~al.}(2020)\citenamefont {Guha},
  \citenamefont {Alimuddin},\ and\ \citenamefont {Parashar}}]{Guha20}%
  \BibitemOpen
  \bibfield  {author} {\bibinfo {author} {\bibfnamefont {Tamal}\ \bibnamefont
  {Guha}}, \bibinfo {author} {\bibfnamefont {Mir}\ \bibnamefont {Alimuddin}}, \
  and\ \bibinfo {author} {\bibfnamefont {Preeti}\ \bibnamefont {Parashar}},\
  }\bibfield  {title} {\enquote {\bibinfo {title} {Thermodynamic advancement in
  the causally inseparable occurrence of thermal maps},}\ }\href@noop {}
  {\bibfield  {journal} {\bibinfo  {journal} {Phys. Rev. A}\ }\textbf {\bibinfo
  {volume} {102}},\ \bibinfo {pages} {032215} (\bibinfo {year}
  {2020})}\BibitemShut {NoStop}%
\bibitem [{\citenamefont {Procopio}\ \emph {et~al.}(2015)\citenamefont
  {Procopio}, \citenamefont {Moqanaki}, \citenamefont {Ara{\'u}jo},
  \citenamefont {Costa}, \citenamefont {Calafell}, \citenamefont {Dowd},
  \citenamefont {Hamel}, \citenamefont {Rozema}, \citenamefont {Brukner},\ and\
  \citenamefont {Walther}}]{Procopio15}%
  \BibitemOpen
  \bibfield  {author} {\bibinfo {author} {\bibfnamefont {Lorenzo~M}\
  \bibnamefont {Procopio}}, \bibinfo {author} {\bibfnamefont {Amir}\
  \bibnamefont {Moqanaki}}, \bibinfo {author} {\bibfnamefont {Mateus}\
  \bibnamefont {Ara{\'u}jo}}, \bibinfo {author} {\bibfnamefont {Fabio}\
  \bibnamefont {Costa}}, \bibinfo {author} {\bibfnamefont {Irati~Alonso}\
  \bibnamefont {Calafell}}, \bibinfo {author} {\bibfnamefont {Emma~G}\
  \bibnamefont {Dowd}}, \bibinfo {author} {\bibfnamefont {Deny~R}\ \bibnamefont
  {Hamel}}, \bibinfo {author} {\bibfnamefont {Lee~A}\ \bibnamefont {Rozema}},
  \bibinfo {author} {\bibfnamefont {{\v{C}}aslav}\ \bibnamefont {Brukner}}, \
  and\ \bibinfo {author} {\bibfnamefont {Philip}\ \bibnamefont {Walther}},\
  }\bibfield  {title} {\enquote {\bibinfo {title} {Experimental superposition
  of orders of quantum gates},}\ }\href@noop {} {\bibfield  {journal} {\bibinfo
   {journal} {Nature Communications}\ }\textbf {\bibinfo {volume} {6}},\
  \bibinfo {pages} {1--6} (\bibinfo {year} {2015})}\BibitemShut {NoStop}%
\bibitem [{\citenamefont {Rubino}\ \emph {et~al.}(2017)\citenamefont {Rubino},
  \citenamefont {Rozema}, \citenamefont {Feix}, \citenamefont {Ara{\'u}jo},
  \citenamefont {Zeuner}, \citenamefont {Procopio}, \citenamefont {Brukner},\
  and\ \citenamefont {Walther}}]{Rubino17}%
  \BibitemOpen
  \bibfield  {author} {\bibinfo {author} {\bibfnamefont {Giulia}\ \bibnamefont
  {Rubino}}, \bibinfo {author} {\bibfnamefont {Lee~A}\ \bibnamefont {Rozema}},
  \bibinfo {author} {\bibfnamefont {Adrien}\ \bibnamefont {Feix}}, \bibinfo
  {author} {\bibfnamefont {Mateus}\ \bibnamefont {Ara{\'u}jo}}, \bibinfo
  {author} {\bibfnamefont {Jonas~M}\ \bibnamefont {Zeuner}}, \bibinfo {author}
  {\bibfnamefont {Lorenzo~M}\ \bibnamefont {Procopio}}, \bibinfo {author}
  {\bibfnamefont {{\v{C}}aslav}\ \bibnamefont {Brukner}}, \ and\ \bibinfo
  {author} {\bibfnamefont {Philip}\ \bibnamefont {Walther}},\ }\bibfield
  {title} {\enquote {\bibinfo {title} {Experimental verification of an
  indefinite causal order},}\ }\href@noop {} {\bibfield  {journal} {\bibinfo
  {journal} {Science Advances}\ }\textbf {\bibinfo {volume} {3}},\ \bibinfo
  {pages} {e1602589} (\bibinfo {year} {2017})}\BibitemShut {NoStop}%
\bibitem [{\citenamefont {Goswami}\ \emph {et~al.}(2018)\citenamefont
  {Goswami}, \citenamefont {Giarmatzi}, \citenamefont {Kewming}, \citenamefont
  {Costa}, \citenamefont {Branciard}, \citenamefont {Romero},\ and\
  \citenamefont {White}}]{Goswami18(1)}%
  \BibitemOpen
  \bibfield  {author} {\bibinfo {author} {\bibfnamefont {K}~\bibnamefont
  {Goswami}}, \bibinfo {author} {\bibfnamefont {Christina}\ \bibnamefont
  {Giarmatzi}}, \bibinfo {author} {\bibfnamefont {M}~\bibnamefont {Kewming}},
  \bibinfo {author} {\bibfnamefont {Fabio}\ \bibnamefont {Costa}}, \bibinfo
  {author} {\bibfnamefont {Cyril}\ \bibnamefont {Branciard}}, \bibinfo {author}
  {\bibfnamefont {Jacquiline}\ \bibnamefont {Romero}}, \ and\ \bibinfo {author}
  {\bibfnamefont {AG}~\bibnamefont {White}},\ }\bibfield  {title} {\enquote
  {\bibinfo {title} {Indefinite causal order in a quantum switch},}\
  }\href@noop {} {\bibfield  {journal} {\bibinfo  {journal} {Physical Review
  Letters}\ }\textbf {\bibinfo {volume} {121}},\ \bibinfo {pages} {090503}
  (\bibinfo {year} {2018})}\BibitemShut {NoStop}%
\bibitem [{\citenamefont {Wei}\ \emph {et~al.}(2019)\citenamefont {Wei},
  \citenamefont {Tischler}, \citenamefont {Zhao}, \citenamefont {Li},
  \citenamefont {Arrazola}, \citenamefont {Liu}, \citenamefont {Zhang},
  \citenamefont {Li}, \citenamefont {You}, \citenamefont {Wang} \emph
  {et~al.}}]{wei2019experimental}%
  \BibitemOpen
  \bibfield  {author} {\bibinfo {author} {\bibfnamefont {Kejin}\ \bibnamefont
  {Wei}}, \bibinfo {author} {\bibfnamefont {Nora}\ \bibnamefont {Tischler}},
  \bibinfo {author} {\bibfnamefont {Si-Ran}\ \bibnamefont {Zhao}}, \bibinfo
  {author} {\bibfnamefont {Yu-Huai}\ \bibnamefont {Li}}, \bibinfo {author}
  {\bibfnamefont {Juan~Miguel}\ \bibnamefont {Arrazola}}, \bibinfo {author}
  {\bibfnamefont {Yang}\ \bibnamefont {Liu}}, \bibinfo {author} {\bibfnamefont
  {Weijun}\ \bibnamefont {Zhang}}, \bibinfo {author} {\bibfnamefont {Hao}\
  \bibnamefont {Li}}, \bibinfo {author} {\bibfnamefont {Lixing}\ \bibnamefont
  {You}}, \bibinfo {author} {\bibfnamefont {Zhen}\ \bibnamefont {Wang}},  \emph
  {et~al.},\ }\bibfield  {title} {\enquote {\bibinfo {title} {Experimental
  quantum switching for exponentially superior quantum communication
  complexity},}\ }\href@noop {} {\bibfield  {journal} {\bibinfo  {journal}
  {Physical Review Letters}\ }\textbf {\bibinfo {volume} {122}},\ \bibinfo
  {pages} {120504} (\bibinfo {year} {2019})}\BibitemShut {NoStop}%
\bibitem [{\citenamefont {Guo}\ \emph {et~al.}(2020)\citenamefont {Guo},
  \citenamefont {Hu}, \citenamefont {Hou}, \citenamefont {Cao}, \citenamefont
  {Cui}, \citenamefont {Liu}, \citenamefont {Huang}, \citenamefont {Li},
  \citenamefont {Guo},\ and\ \citenamefont {Chiribella}}]{guo2020experimental}%
  \BibitemOpen
  \bibfield  {author} {\bibinfo {author} {\bibfnamefont {Yu}~\bibnamefont
  {Guo}}, \bibinfo {author} {\bibfnamefont {Xiao-Min}\ \bibnamefont {Hu}},
  \bibinfo {author} {\bibfnamefont {Zhi-Bo}\ \bibnamefont {Hou}}, \bibinfo
  {author} {\bibfnamefont {Huan}\ \bibnamefont {Cao}}, \bibinfo {author}
  {\bibfnamefont {Jin-Ming}\ \bibnamefont {Cui}}, \bibinfo {author}
  {\bibfnamefont {Bi-Heng}\ \bibnamefont {Liu}}, \bibinfo {author}
  {\bibfnamefont {Yun-Feng}\ \bibnamefont {Huang}}, \bibinfo {author}
  {\bibfnamefont {Chuan-Feng}\ \bibnamefont {Li}}, \bibinfo {author}
  {\bibfnamefont {Guang-Can}\ \bibnamefont {Guo}}, \ and\ \bibinfo {author}
  {\bibfnamefont {Giulio}\ \bibnamefont {Chiribella}},\ }\bibfield  {title}
  {\enquote {\bibinfo {title} {Experimental transmission of quantum information
  using a superposition of causal orders},}\ }\href@noop {} {\bibfield
  {journal} {\bibinfo  {journal} {Physical Review Letters}\ }\textbf {\bibinfo
  {volume} {124}},\ \bibinfo {pages} {030502} (\bibinfo {year}
  {2020})}\BibitemShut {NoStop}%
\bibitem [{\citenamefont {Goswami}\ and\ \citenamefont
  {Romero}(2020)}]{goswami2020experiments}%
  \BibitemOpen
  \bibfield  {author} {\bibinfo {author} {\bibfnamefont {K}~\bibnamefont
  {Goswami}}\ and\ \bibinfo {author} {\bibfnamefont {J}~\bibnamefont
  {Romero}},\ }\bibfield  {title} {\enquote {\bibinfo {title} {Experiments on
  quantum causality},}\ }\href@noop {} {\bibfield  {journal} {\bibinfo
  {journal} {AVS Quantum Science}\ }\textbf {\bibinfo {volume} {2}},\ \bibinfo
  {pages} {037101} (\bibinfo {year} {2020})}\BibitemShut {NoStop}%
\bibitem [{\citenamefont {Rubino}\ \emph {et~al.}(2020)\citenamefont {Rubino},
  \citenamefont {Rozema}, \citenamefont {Ebler}, \citenamefont
  {Kristj{\'a}nsson}, \citenamefont {Salek}, \citenamefont {Gu{\'e}rin},
  \citenamefont {Abbott}, \citenamefont {Branciard}, \citenamefont {Brukner},
  \citenamefont {Chiribella} \emph {et~al.}}]{rubino2020experimental}%
  \BibitemOpen
  \bibfield  {author} {\bibinfo {author} {\bibfnamefont {Giulia}\ \bibnamefont
  {Rubino}}, \bibinfo {author} {\bibfnamefont {Lee~A}\ \bibnamefont {Rozema}},
  \bibinfo {author} {\bibfnamefont {Daniel}\ \bibnamefont {Ebler}}, \bibinfo
  {author} {\bibfnamefont {Hl{\'e}r}\ \bibnamefont {Kristj{\'a}nsson}},
  \bibinfo {author} {\bibfnamefont {Sina}\ \bibnamefont {Salek}}, \bibinfo
  {author} {\bibfnamefont {Philippe~Allard}\ \bibnamefont {Gu{\'e}rin}},
  \bibinfo {author} {\bibfnamefont {Alastair~A}\ \bibnamefont {Abbott}},
  \bibinfo {author} {\bibfnamefont {Cyril}\ \bibnamefont {Branciard}}, \bibinfo
  {author} {\bibfnamefont {{\v{C}}aslav}\ \bibnamefont {Brukner}}, \bibinfo
  {author} {\bibfnamefont {Giulio}\ \bibnamefont {Chiribella}},  \emph
  {et~al.},\ }\bibfield  {title} {\enquote {\bibinfo {title} {Experimental
  quantum communication enhancement by superposing trajectories},}\ }\href@noop
  {} {\bibfield  {journal} {\bibinfo  {journal} {arXiv preprint
  arXiv:2007.05005}\ } (\bibinfo {year} {2020})}\BibitemShut {NoStop}%
\bibitem [{\citenamefont {Zych}\ \emph {et~al.}(2019)\citenamefont {Zych},
  \citenamefont {Costa}, \citenamefont {Pikovski},\ and\ \citenamefont
  {Brukner}}]{Zych17}%
  \BibitemOpen
  \bibfield  {author} {\bibinfo {author} {\bibfnamefont {Magdalena}\
  \bibnamefont {Zych}}, \bibinfo {author} {\bibfnamefont {Fabio}\ \bibnamefont
  {Costa}}, \bibinfo {author} {\bibfnamefont {Igor}\ \bibnamefont {Pikovski}},
  \ and\ \bibinfo {author} {\bibfnamefont {{\v{C}}aslav}\ \bibnamefont
  {Brukner}},\ }\bibfield  {title} {\enquote {\bibinfo {title} {Bell’s
  theorem for temporal order},}\ }\href@noop {} {\bibfield  {journal} {\bibinfo
   {journal} {Nature Communications}\ }\textbf {\bibinfo {volume} {10}},\
  \bibinfo {pages} {1--10} (\bibinfo {year} {2019})}\BibitemShut {NoStop}%
\bibitem [{\citenamefont {Ebler}\ \emph {et~al.}(2018)\citenamefont {Ebler},
  \citenamefont {Salek},\ and\ \citenamefont {Chiribella}}]{Ebler18}%
  \BibitemOpen
  \bibfield  {author} {\bibinfo {author} {\bibfnamefont {Daniel}\ \bibnamefont
  {Ebler}}, \bibinfo {author} {\bibfnamefont {Sina}\ \bibnamefont {Salek}}, \
  and\ \bibinfo {author} {\bibfnamefont {Giulio}\ \bibnamefont {Chiribella}},\
  }\bibfield  {title} {\enquote {\bibinfo {title} {Enhanced communication with
  the assistance of indefinite causal order},}\ }\href@noop {} {\bibfield
  {journal} {\bibinfo  {journal} {Physical Review Letters}\ }\textbf {\bibinfo
  {volume} {120}},\ \bibinfo {pages} {120502} (\bibinfo {year}
  {2018})}\BibitemShut {NoStop}%
\bibitem [{\citenamefont {Salek}\ \emph {et~al.}(2018)\citenamefont {Salek},
  \citenamefont {Ebler},\ and\ \citenamefont {Chiribella}}]{Salek18}%
  \BibitemOpen
  \bibfield  {author} {\bibinfo {author} {\bibfnamefont {Sina}\ \bibnamefont
  {Salek}}, \bibinfo {author} {\bibfnamefont {Daniel}\ \bibnamefont {Ebler}}, \
  and\ \bibinfo {author} {\bibfnamefont {Giulio}\ \bibnamefont {Chiribella}},\
  }\bibfield  {title} {\enquote {\bibinfo {title} {Quantum communication in a
  superposition of causal orders},}\ }\href@noop {} {\bibfield  {journal}
  {\bibinfo  {journal} {arXiv preprint arXiv:1809.06655}\ } (\bibinfo {year}
  {2018})}\BibitemShut {NoStop}%
\bibitem [{\citenamefont {Kristjánsson}\ \emph {et~al.}(2020)\citenamefont
  {Kristjánsson}, \citenamefont {Chiribella}, \citenamefont {Salek},
  \citenamefont {Ebler},\ and\ \citenamefont {Wilson}}]{Hler20}%
  \BibitemOpen
  \bibfield  {author} {\bibinfo {author} {\bibfnamefont {H.}~\bibnamefont
  {Kristjánsson}}, \bibinfo {author} {\bibfnamefont {G.}~\bibnamefont
  {Chiribella}}, \bibinfo {author} {\bibfnamefont {S.}~\bibnamefont {Salek}},
  \bibinfo {author} {\bibfnamefont {D.}~\bibnamefont {Ebler}}, \ and\ \bibinfo
  {author} {\bibfnamefont {M.}~\bibnamefont {Wilson}},\ }\bibfield  {title}
  {\enquote {\bibinfo {title} {Resource theories of communication},}\
  }\href@noop {} {\bibfield  {journal} {\bibinfo  {journal} {New J. Phys.}\
  }\textbf {\bibinfo {volume} {22}},\ \bibinfo {pages} {073014} (\bibinfo
  {year} {2020})}\BibitemShut {NoStop}%
\bibitem [{\citenamefont {Goswami}\ \emph {et~al.}(2020)\citenamefont
  {Goswami}, \citenamefont {Cao}, \citenamefont {Paz-Silva}, \citenamefont
  {Romero},\ and\ \citenamefont {White}}]{Goswami18}%
  \BibitemOpen
  \bibfield  {author} {\bibinfo {author} {\bibfnamefont {K.}~\bibnamefont
  {Goswami}}, \bibinfo {author} {\bibfnamefont {Y.}~\bibnamefont {Cao}},
  \bibinfo {author} {\bibfnamefont {G.~A.}\ \bibnamefont {Paz-Silva}}, \bibinfo
  {author} {\bibfnamefont {J.}~\bibnamefont {Romero}}, \ and\ \bibinfo {author}
  {\bibfnamefont {A.~G.}\ \bibnamefont {White}},\ }\bibfield  {title} {\enquote
  {\bibinfo {title} {Increasing communication capacity via superposition of
  order},}\ }\href@noop {} {\bibfield  {journal} {\bibinfo  {journal} {Phys.
  Rev. Research}\ }\textbf {\bibinfo {volume} {2}},\ \bibinfo {pages} {033292}
  (\bibinfo {year} {2020})}\BibitemShut {NoStop}%
\bibitem [{\citenamefont {Abbott}\ \emph {et~al.}(2020)\citenamefont {Abbott},
  \citenamefont {Wechs}, \citenamefont {Horsman}, \citenamefont {Mhalla},\ and\
  \citenamefont {Branciard}}]{Abbott2018}%
  \BibitemOpen
  \bibfield  {author} {\bibinfo {author} {\bibfnamefont {Alastair~A}\
  \bibnamefont {Abbott}}, \bibinfo {author} {\bibfnamefont {Julian}\
  \bibnamefont {Wechs}}, \bibinfo {author} {\bibfnamefont {Dominic}\
  \bibnamefont {Horsman}}, \bibinfo {author} {\bibfnamefont {Mehdi}\
  \bibnamefont {Mhalla}}, \ and\ \bibinfo {author} {\bibfnamefont {Cyril}\
  \bibnamefont {Branciard}},\ }\bibfield  {title} {\enquote {\bibinfo {title}
  {Communication through coherent control of quantum channels},}\ }\href@noop
  {} {\bibfield  {journal} {\bibinfo  {journal} {Quantum}\ }\textbf {\bibinfo
  {volume} {4}},\ \bibinfo {pages} {333} (\bibinfo {year} {2020})}\BibitemShut
  {NoStop}%
\bibitem [{\citenamefont {Chiribella}\ and\ \citenamefont
  {Kristj{\'a}nsson}(2019{\natexlab{a}})}]{chiribella2019quantum}%
  \BibitemOpen
  \bibfield  {author} {\bibinfo {author} {\bibfnamefont {Giulio}\ \bibnamefont
  {Chiribella}}\ and\ \bibinfo {author} {\bibfnamefont {Hl{\'e}r}\ \bibnamefont
  {Kristj{\'a}nsson}},\ }\bibfield  {title} {\enquote {\bibinfo {title}
  {Quantum shannon theory with superpositions of trajectories},}\ }\href@noop
  {} {\bibfield  {journal} {\bibinfo  {journal} {Proceedings of the Royal
  Society A}\ }\textbf {\bibinfo {volume} {475}},\ \bibinfo {pages} {20180903}
  (\bibinfo {year} {2019}{\natexlab{a}})}\BibitemShut {NoStop}%
\bibitem [{\citenamefont {Kristj{\'a}nsson}\ \emph {et~al.}(2020)\citenamefont
  {Kristj{\'a}nsson}, \citenamefont {Mao},\ and\ \citenamefont
  {Chiribella}}]{kristjansson2020single}%
  \BibitemOpen
  \bibfield  {author} {\bibinfo {author} {\bibfnamefont {Hl{\'e}r}\
  \bibnamefont {Kristj{\'a}nsson}}, \bibinfo {author} {\bibfnamefont {Wenxu}\
  \bibnamefont {Mao}}, \ and\ \bibinfo {author} {\bibfnamefont {Giulio}\
  \bibnamefont {Chiribella}},\ }\bibfield  {title} {\enquote {\bibinfo {title}
  {Single-particle communication through correlated noise},}\ }\href@noop {}
  {\bibfield  {journal} {\bibinfo  {journal} {arXiv preprint arXiv:2004.06090}\
  } (\bibinfo {year} {2020})}\BibitemShut {NoStop}%
\bibitem [{\citenamefont {Chiribella}\ \emph {et~al.}(2008)\citenamefont
  {Chiribella}, \citenamefont {D'Ariano},\ and\ \citenamefont
  {Perinotti}}]{chiribella2008transforming}%
  \BibitemOpen
  \bibfield  {author} {\bibinfo {author} {\bibfnamefont {Giulio}\ \bibnamefont
  {Chiribella}}, \bibinfo {author} {\bibfnamefont {Giacomo~Mauro}\ \bibnamefont
  {D'Ariano}}, \ and\ \bibinfo {author} {\bibfnamefont {Paolo}\ \bibnamefont
  {Perinotti}},\ }\bibfield  {title} {\enquote {\bibinfo {title} {Transforming
  quantum operations: Quantum supermaps},}\ }\href@noop {} {\bibfield
  {journal} {\bibinfo  {journal} {EPL (Europhysics Letters)}\ }\textbf
  {\bibinfo {volume} {83}},\ \bibinfo {pages} {30004} (\bibinfo {year}
  {2008})}\BibitemShut {NoStop}%
\bibitem [{\citenamefont {Chiribella}\ \emph
  {et~al.}(2009{\natexlab{b}})\citenamefont {Chiribella}, \citenamefont
  {D’Ariano},\ and\ \citenamefont {Perinotti}}]{chiribella2009theoretical}%
  \BibitemOpen
  \bibfield  {author} {\bibinfo {author} {\bibfnamefont {Giulio}\ \bibnamefont
  {Chiribella}}, \bibinfo {author} {\bibfnamefont {Giacomo~Mauro}\ \bibnamefont
  {D’Ariano}}, \ and\ \bibinfo {author} {\bibfnamefont {Paolo}\ \bibnamefont
  {Perinotti}},\ }\bibfield  {title} {\enquote {\bibinfo {title} {Theoretical
  framework for quantum networks},}\ }\href@noop {} {\bibfield  {journal}
  {\bibinfo  {journal} {Physical Review A}\ }\textbf {\bibinfo {volume} {80}},\
  \bibinfo {pages} {022339} (\bibinfo {year} {2009}{\natexlab{b}})}\BibitemShut
  {NoStop}%
\bibitem [{\citenamefont {Gregoratti}\ and\ \citenamefont
  {Werner}(2003)}]{Greg03}%
  \BibitemOpen
  \bibfield  {author} {\bibinfo {author} {\bibfnamefont {Matteo}\ \bibnamefont
  {Gregoratti}}\ and\ \bibinfo {author} {\bibfnamefont {Reinhard~F}\
  \bibnamefont {Werner}},\ }\bibfield  {title} {\enquote {\bibinfo {title}
  {Quantum lost and found},}\ }\href@noop {} {\bibfield  {journal} {\bibinfo
  {journal} {Journal of Modern Optics}\ }\textbf {\bibinfo {volume} {50}},\
  \bibinfo {pages} {915--933} (\bibinfo {year} {2003})}\BibitemShut {NoStop}%
\bibitem [{\citenamefont {Smolin}\ \emph {et~al.}(2005)\citenamefont {Smolin},
  \citenamefont {Verstraete},\ and\ \citenamefont {Winter}}]{Smolin05}%
  \BibitemOpen
  \bibfield  {author} {\bibinfo {author} {\bibfnamefont {John~A}\ \bibnamefont
  {Smolin}}, \bibinfo {author} {\bibfnamefont {Frank}\ \bibnamefont
  {Verstraete}}, \ and\ \bibinfo {author} {\bibfnamefont {Andreas}\
  \bibnamefont {Winter}},\ }\bibfield  {title} {\enquote {\bibinfo {title}
  {Entanglement of assistance and multipartite state distillation},}\
  }\href@noop {} {\bibfield  {journal} {\bibinfo  {journal} {Physical Review
  A}\ }\textbf {\bibinfo {volume} {72}},\ \bibinfo {pages} {052317} (\bibinfo
  {year} {2005})}\BibitemShut {NoStop}%
\bibitem [{\citenamefont {Schumacher}\ and\ \citenamefont
  {Nielsen}(1996)}]{Schumacher96}%
  \BibitemOpen
  \bibfield  {author} {\bibinfo {author} {\bibfnamefont {Benjamin}\
  \bibnamefont {Schumacher}}\ and\ \bibinfo {author} {\bibfnamefont
  {Michael~A}\ \bibnamefont {Nielsen}},\ }\bibfield  {title} {\enquote
  {\bibinfo {title} {Quantum data processing and error correction},}\
  }\href@noop {} {\bibfield  {journal} {\bibinfo  {journal} {Physical Review
  A}\ }\textbf {\bibinfo {volume} {54}},\ \bibinfo {pages} {2629} (\bibinfo
  {year} {1996})}\BibitemShut {NoStop}%
\bibitem [{\citenamefont {Holevo}\ and\ \citenamefont
  {Werner}(2001)}]{Holevo99}%
  \BibitemOpen
  \bibfield  {author} {\bibinfo {author} {\bibfnamefont {Alexander~S}\
  \bibnamefont {Holevo}}\ and\ \bibinfo {author} {\bibfnamefont {Reinhard~F}\
  \bibnamefont {Werner}},\ }\bibfield  {title} {\enquote {\bibinfo {title}
  {Evaluating capacities of bosonic gaussian channels},}\ }\href@noop {}
  {\bibfield  {journal} {\bibinfo  {journal} {Physical Review A}\ }\textbf
  {\bibinfo {volume} {63}},\ \bibinfo {pages} {032312} (\bibinfo {year}
  {2001})}\BibitemShut {NoStop}%
\bibitem [{\citenamefont {Holevo}(2008)}]{Holevo08}%
  \BibitemOpen
  \bibfield  {author} {\bibinfo {author} {\bibfnamefont {Alexander~S}\
  \bibnamefont {Holevo}},\ }\bibfield  {title} {\enquote {\bibinfo {title}
  {Entanglement-breaking channels in infinite dimensions},}\ }\href@noop {}
  {\bibfield  {journal} {\bibinfo  {journal} {Problems of Information
  Transmission}\ }\textbf {\bibinfo {volume} {44}},\ \bibinfo {pages}
  {171--184} (\bibinfo {year} {2008})}\BibitemShut {NoStop}%
\bibitem [{\citenamefont {Streltsov}\ \emph {et~al.}(2017)\citenamefont
  {Streltsov}, \citenamefont {Adesso},\ and\ \citenamefont
  {Plenio}}]{streltsov2017colloquium}%
  \BibitemOpen
  \bibfield  {author} {\bibinfo {author} {\bibfnamefont {Alexander}\
  \bibnamefont {Streltsov}}, \bibinfo {author} {\bibfnamefont {Gerardo}\
  \bibnamefont {Adesso}}, \ and\ \bibinfo {author} {\bibfnamefont {Martin~B}\
  \bibnamefont {Plenio}},\ }\bibfield  {title} {\enquote {\bibinfo {title}
  {Colloquium: Quantum coherence as a resource},}\ }\href@noop {} {\bibfield
  {journal} {\bibinfo  {journal} {Reviews of Modern Physics}\ }\textbf
  {\bibinfo {volume} {89}},\ \bibinfo {pages} {041003} (\bibinfo {year}
  {2017})}\BibitemShut {NoStop}%
\bibitem [{\citenamefont {Devetak}\ and\ \citenamefont
  {Winter}(2005)}]{Devetak05(1)}%
  \BibitemOpen
  \bibfield  {author} {\bibinfo {author} {\bibfnamefont {Igor}\ \bibnamefont
  {Devetak}}\ and\ \bibinfo {author} {\bibfnamefont {Andreas}\ \bibnamefont
  {Winter}},\ }\bibfield  {title} {\enquote {\bibinfo {title} {Distillation of
  secret key and entanglement from quantum states},}\ }\href@noop {} {\bibfield
   {journal} {\bibinfo  {journal} {Proceedings of the Royal Society A:
  Mathematical, Physical and engineering sciences}\ }\textbf {\bibinfo {volume}
  {461}},\ \bibinfo {pages} {207--235} (\bibinfo {year} {2005})}\BibitemShut
  {NoStop}%
\bibitem [{\citenamefont {Colnaghi}\ \emph {et~al.}(2012)\citenamefont
  {Colnaghi}, \citenamefont {D'Ariano}, \citenamefont {Facchini},\ and\
  \citenamefont {Perinotti}}]{colnaghi2012quantum}%
  \BibitemOpen
  \bibfield  {author} {\bibinfo {author} {\bibfnamefont {Timoteo}\ \bibnamefont
  {Colnaghi}}, \bibinfo {author} {\bibfnamefont {Giacomo~Mauro}\ \bibnamefont
  {D'Ariano}}, \bibinfo {author} {\bibfnamefont {Stefano}\ \bibnamefont
  {Facchini}}, \ and\ \bibinfo {author} {\bibfnamefont {Paolo}\ \bibnamefont
  {Perinotti}},\ }\bibfield  {title} {\enquote {\bibinfo {title} {Quantum
  computation with programmable connections between gates},}\ }\href@noop {}
  {\bibfield  {journal} {\bibinfo  {journal} {Physics Letters A}\ }\textbf
  {\bibinfo {volume} {376}},\ \bibinfo {pages} {2940--2943} (\bibinfo {year}
  {2012})}\BibitemShut {NoStop}%
\bibitem [{\citenamefont {Facchini}\ and\ \citenamefont
  {Perdrix}(2015)}]{facchini2015quantum}%
  \BibitemOpen
  \bibfield  {author} {\bibinfo {author} {\bibfnamefont {Stefano}\ \bibnamefont
  {Facchini}}\ and\ \bibinfo {author} {\bibfnamefont {Simon}\ \bibnamefont
  {Perdrix}},\ }\bibfield  {title} {\enquote {\bibinfo {title} {Quantum
  circuits for the unitary permutation problem},}\ }in\ \href@noop {} {\emph
  {\bibinfo {booktitle} {International Conference on Theory and Applications of
  Models of Computation}}}\ (\bibinfo {organization} {Springer},\ \bibinfo
  {year} {2015})\ pp.\ \bibinfo {pages} {324--331}\BibitemShut {NoStop}%
\bibitem [{\citenamefont {Chiribella}\ and\ \citenamefont
  {Kristj{\'a}nsson}(2019{\natexlab{b}})}]{Giulio18}%
  \BibitemOpen
  \bibfield  {author} {\bibinfo {author} {\bibfnamefont {Giulio}\ \bibnamefont
  {Chiribella}}\ and\ \bibinfo {author} {\bibfnamefont {Hl{\'e}r}\ \bibnamefont
  {Kristj{\'a}nsson}},\ }\bibfield  {title} {\enquote {\bibinfo {title}
  {Quantum shannon theory with superpositions of trajectories},}\ }\href@noop
  {} {\bibfield  {journal} {\bibinfo  {journal} {Proceedings of the Royal
  Society A : Mathematical, Physical and engineering sciences}\ }\textbf
  {\bibinfo {volume} {475}},\ \bibinfo {pages} {20180903} (\bibinfo {year}
  {2019}{\natexlab{b}})}\BibitemShut {NoStop}%
\bibitem [{\citenamefont {Chiribella}\ and\ \citenamefont
  {Yang}(2017)}]{chiribella2017optimal}%
  \BibitemOpen
  \bibfield  {author} {\bibinfo {author} {\bibfnamefont {Giulio}\ \bibnamefont
  {Chiribella}}\ and\ \bibinfo {author} {\bibfnamefont {Yuxiang}\ \bibnamefont
  {Yang}},\ }\bibfield  {title} {\enquote {\bibinfo {title} {Optimal quantum
  operations at zero energy cost},}\ }\href@noop {} {\bibfield  {journal}
  {\bibinfo  {journal} {Physical Review A}\ }\textbf {\bibinfo {volume} {96}},\
  \bibinfo {pages} {022327} (\bibinfo {year} {2017})}\BibitemShut {NoStop}%
\bibitem [{\citenamefont {Oreshkov}(2019)}]{Oreshkov2018}%
  \BibitemOpen
  \bibfield  {author} {\bibinfo {author} {\bibfnamefont {Ognyan}\ \bibnamefont
  {Oreshkov}},\ }\bibfield  {title} {\enquote {\bibinfo {title}
  {Time-delocalized quantum subsystems and operations: on the existence of
  processes with indefinite causal structure in quantum mechanics},}\
  }\href@noop {} {\bibfield  {journal} {\bibinfo  {journal} {Quantum}\ }\textbf
  {\bibinfo {volume} {3}},\ \bibinfo {pages} {206} (\bibinfo {year}
  {2019})}\BibitemShut {NoStop}%
\bibitem [{\citenamefont {Ziman}\ and\ \citenamefont
  {Bu{\v{z}}ek}(2005)}]{Ziman05}%
  \BibitemOpen
  \bibfield  {author} {\bibinfo {author} {\bibfnamefont {M{\'a}rio}\
  \bibnamefont {Ziman}}\ and\ \bibinfo {author} {\bibfnamefont {Vladim{\'\i}r}\
  \bibnamefont {Bu{\v{z}}ek}},\ }\bibfield  {title} {\enquote {\bibinfo {title}
  {All (qubit) decoherences: Complete characterization and physical
  implementation},}\ }\href@noop {} {\bibfield  {journal} {\bibinfo  {journal}
  {Physical Review A}\ }\textbf {\bibinfo {volume} {72}},\ \bibinfo {pages}
  {022110} (\bibinfo {year} {2005})}\BibitemShut {NoStop}%
\bibitem [{\citenamefont {Knill}\ and\ \citenamefont
  {Laflamme}(1997)}]{Knill96}%
  \BibitemOpen
  \bibfield  {author} {\bibinfo {author} {\bibfnamefont {Emanuel}\ \bibnamefont
  {Knill}}\ and\ \bibinfo {author} {\bibfnamefont {Raymond}\ \bibnamefont
  {Laflamme}},\ }\bibfield  {title} {\enquote {\bibinfo {title} {Theory of
  quantum error-correcting codes},}\ }\href@noop {} {\bibfield  {journal}
  {\bibinfo  {journal} {Physical Review A}\ }\textbf {\bibinfo {volume} {55}},\
  \bibinfo {pages} {900} (\bibinfo {year} {1997})}\BibitemShut {NoStop}%
\bibitem [{\citenamefont {Chiribella}\ \emph {et~al.}(2011)\citenamefont
  {Chiribella}, \citenamefont {Dall’Arno}, \citenamefont {D’Ariano},
  \citenamefont {Macchiavello},\ and\ \citenamefont {Perinotti}}]{Giulio11}%
  \BibitemOpen
  \bibfield  {author} {\bibinfo {author} {\bibfnamefont {Giulio}\ \bibnamefont
  {Chiribella}}, \bibinfo {author} {\bibfnamefont {Michele}\ \bibnamefont
  {Dall’Arno}}, \bibinfo {author} {\bibfnamefont {Giacomo~Mauro}\
  \bibnamefont {D’Ariano}}, \bibinfo {author} {\bibfnamefont {Chiara}\
  \bibnamefont {Macchiavello}}, \ and\ \bibinfo {author} {\bibfnamefont
  {Paolo}\ \bibnamefont {Perinotti}},\ }\bibfield  {title} {\enquote {\bibinfo
  {title} {Quantum error correction with degenerate codes for correlated
  noise},}\ }\href@noop {} {\bibfield  {journal} {\bibinfo  {journal} {Physical
  Review A}\ }\textbf {\bibinfo {volume} {83}},\ \bibinfo {pages} {052305}
  (\bibinfo {year} {2011})}\BibitemShut {NoStop}%
\end{thebibliography}%

\end{document}